\documentclass[10pt,conference,compsocconf]{IEEEtran}
\IEEEoverridecommandlockouts
\usepackage{cite}
\usepackage{amsmath,amssymb,amsfonts}
\usepackage{algorithmic}
\usepackage{graphicx}
\usepackage{textcomp}
\usepackage{xcolor}
\usepackage{subscript}

\usepackage{url}
\usepackage{booktabs}
\usepackage{caption}
\usepackage{enumitem}
\usepackage{dsfont}
\usepackage{caption}
\usepackage{subcaption}
\usepackage{xspace}
\usepackage{amsthm}
\usepackage{multirow}
\theoremstyle{definition}

\usepackage[ruled, vlined, linesnumbered]{algorithm2e} 

\def\BibTeX{{\rm B\kern-.05em{\sc i\kern-.025em b}\kern-.08em
    T\kern-.1667em\lower.7ex\hbox{E}\kern-.125emX}}
\begin{document}

\setlength{\textfloatsep}{0.11cm}
\setlength{\dbltextfloatsep}{0.11cm}
\setlength{\abovecaptionskip}{0.11cm}
\setlength{\skip\footins}{0.11cm}

\title{THyMe+: Temporal Hypergraph Motifs and Fast Algorithms for Exact Counting}

\author{
	\IEEEauthorblockN{Geon Lee}
	\IEEEauthorblockA{Graduate School of AI, KAIST\\geonlee0325@kaist.ac.kr}
	\and
	\IEEEauthorblockN{Kijung Shin}
	\IEEEauthorblockA{Graduate School of AI and School of Electrical Engineering, KAIST\\kijungs@kaist.ac.kr}
}

\maketitle
    
\definecolor{myred}{RGB}{195, 79, 82}
\definecolor{mygreen}{RGB}{86, 167 104}
\definecolor{myblue}{RGB}{74, 113 175}

\newcommand\red[1]{\textcolor{red}{#1}}
\newcommand\blue[1]{\textcolor{blue}{#1}}
\newcommand\geon[1]{\textcolor{blue}{[Geon:#1]}}
\newcommand\myred[1]{\textcolor{myred}{#1}}
\newcommand\myblue[1]{\textcolor{myblue}{#1}}
\newcommand\mygreen[1]{\textcolor{mygreen}{#1}}

\newcommand{\short}{TH-motif\xspace}
\newcommand{\shorts}{TH-motifs\xspace}

\newcommand{\motif}{TH-motif\xspace}
\newcommand{\motifs}{TH-motifs\xspace}

\newcommand{\static}{static h-motif\xspace}
\newcommand{\statics}{static h-motifs\xspace}

\newcommand{\smallsection}[1]{{\vspace{0.05in} \noindent {\bf{\underline{\smash{#1}}}}}}
\newtheorem{obs}{\textbf{Observation}}
\newtheorem{dfn}{\textbf{Definition}}
\newtheorem{thm}{\textbf{Theorem}}
\newtheorem{axm}{\textbf{Axiom}}
\newtheorem{lma}{\textbf{Lemma}}
\newtheorem{pro}{\textbf{Problem}}

\newcommand{\cmark}{\ding{51}}%
\newcommand{\xmark}{\ding{55}}%

\newcommand{\wsdm}{\textsc{Dynamic Programming}\xspace}
\newcommand{\wsdmshort}{\textsc{DP}\xspace}
\newcommand{\naive}{\textsc{THyMe}\xspace}
\newcommand{\adv}{\textsc{THyMe\textsuperscript{+}}\xspace}

\newcommand{\triple}{triple-inducing\xspace}
\newcommand{\pair}{pair-inducing\xspace}
\newcommand{\single}{single-inducing\xspace}

\newcommand{\Triple}{Triple-inducing\xspace}
\newcommand{\Pair}{Pair-inducing\xspace}
\newcommand{\Single}{Single-inducing\xspace}

\newcommand{\mybox}{%
    \collectbox{%
        \setlength{\fboxsep}{1pt}%
        \fbox{\BOXCONTENT}%
    }%
}

	\begin{abstract}
Group interactions arise in our daily lives (email communications,  on-demand ride sharing, comment interactions on online communities, to name a few), and they together form hypergraphs that evolve over time.
Given such temporal hypergraphs, how can we describe their underlying design principles?
If their sizes and time spans are considerably different, how can we compare their structural and temporal characteristics?

In this work, we define 96 \textit{temporal hypergraph motifs} (\shorts), 
and propose the relative occurrences of their instances as an answer to the above questions.
\shorts categorize the relational and temporal dynamics among three connected hyperedges that appear within a short time.
For scalable analysis, we develop \adv, a fast and exact algorithm for counting the instances of \shorts in massive hypergraphs, and show that \adv is at most $\mathit{2,163 \times}$ \textit{faster} while requiring less space than baseline.
Using it, we investigate $11$ real-world temporal hypergraphs from various domains.
We demonstrate that \shorts provide important information useful for downstream tasks and reveal interesting patterns, including the striking similarity between temporal hypergraphs from the same domain.






	\end{abstract}
	
	\section{Introduction}
	\label{sec:intro}
	\begin{figure}[t]
	\vspace{-10mm}
\end{figure}

\begin{figure}[t]
	\vspace{-7mm}
	\centering
	\begin{subfigure}[b]{.5\textwidth}
    	\includegraphics[width=0.95\linewidth]{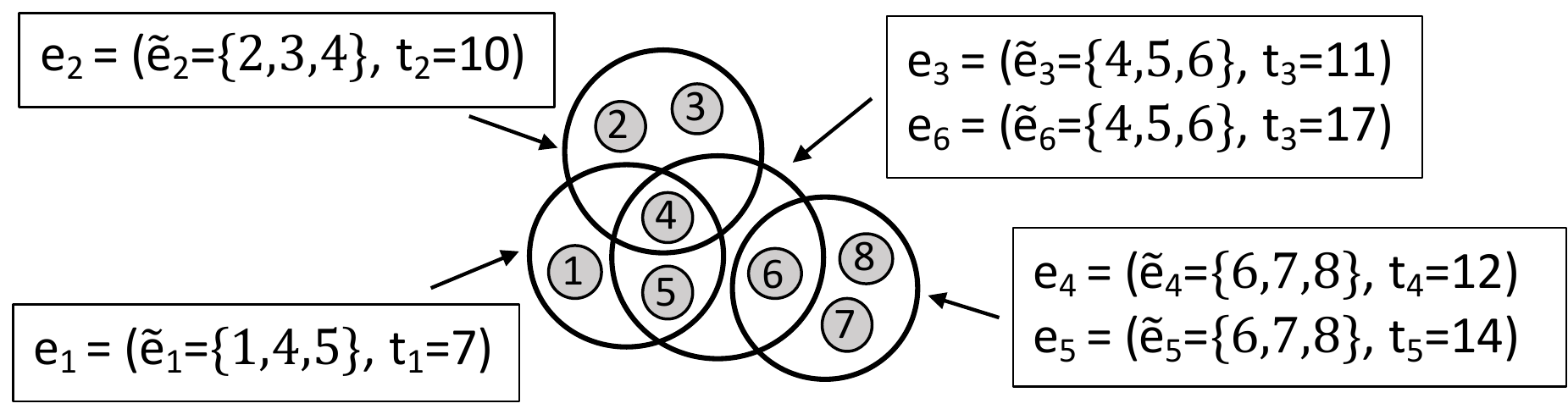}
    	\vspace{-2pt}
    	\caption{An example temporal hypergraph}
	\end{subfigure}
	\begin{subfigure}[b]{.15\textwidth}
    	\includegraphics[width=0.95\linewidth]{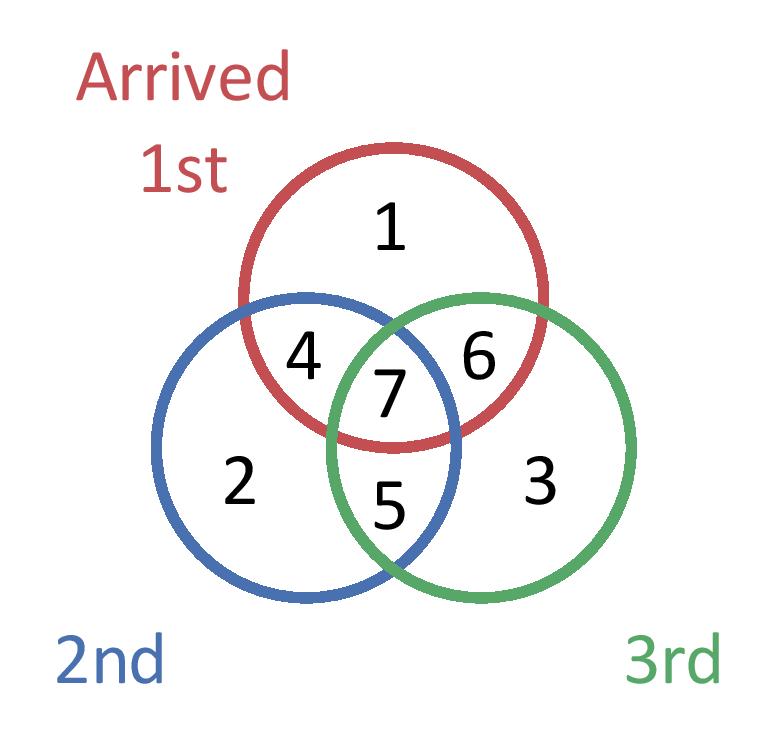}
    	\vspace{-8pt}
    	\caption{7 regions for \\ defining \shorts}
	\end{subfigure}
	\begin{subfigure}[b]{.15\textwidth}
    	\includegraphics[width=0.95\linewidth]{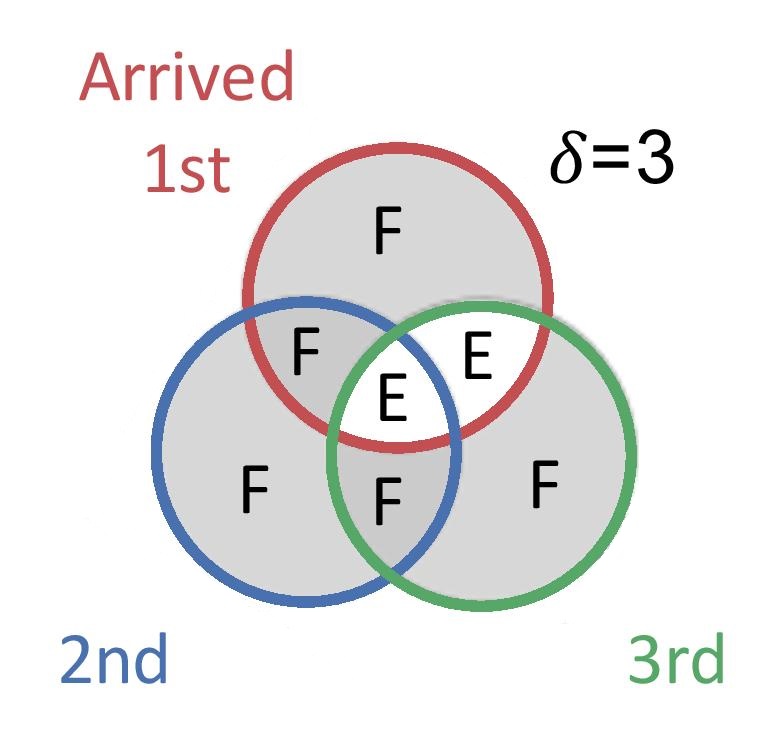}
    	\vspace{-8pt}
    	\caption{The definition \\ $~$ of \motif 77}
	\end{subfigure}
	\begin{subfigure}[b]{.15\textwidth}
    	\includegraphics[width=0.95\linewidth]{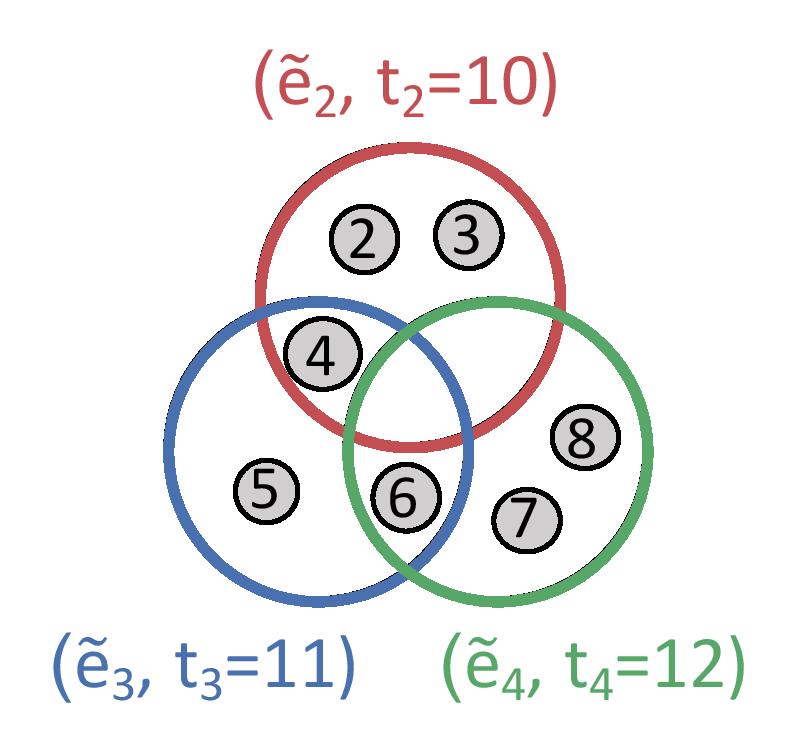}
    	\vspace{-8pt}
    	\caption{An instance  \\ of \motif 77}
	\end{subfigure}
	\caption{\textbf{(a)} A temporal hypergraph with $8$ nodes and $6$ temporal hyperedges. \textbf{(b)} The $7$ regions in the Venn diagram representation  for defining \motifs. \textbf{(c)} The definition of \motif 77. `F' and `E' stand for `filled' and `empty', respectively. \textbf{(d)} The sequence $\langle e_2,e_3,e_4 \rangle$ is an instance of \motif 77.\label{fig:example}}
\end{figure}

Interactions in real-world systems are complex, and in many cases, they are beyond pairwise: email communications,  on-demand ride sharing, comment interactions on online communities, to name a few.
These group interactions together form a \textit{hypergraph}, which consists of a set of nodes and a set of hyperedges (see Fig.~\ref{fig:example}(a) for an example). 
Each \textit{hyperedge} is a subset of \textit{any} number of nodes, and by naturally representing a group interaction among multiple individuals or objects, it contributes to the powerful expressiveness of hypergraphs.

Recently, several empirical studies have revealed structural and temporal properties of real-world hypergraphs.
Pervasive structural patterns include (a) heavy-tailed distributions of degrees, edge sizes, and intersection sizes \cite{kook2020evolution}; (b) giant connected components \cite{do2020structural}, and small diameters \cite{do2020structural}; and (c) substantial overlaps of hyperedges with homophily \cite{lee2021hyperedges}.
Temporal properties observed commonly in various time-evolving hypergraphs include (a) significant overlaps between temporally adjacent hyperedges~\cite{benson2018sequences}; and (b) diminishing overlaps, densification, and shrinking diameters~\cite{kook2020evolution}.

In addition to these macroscopic properties, local connectivity and dynamics in real-world hypergraphs have been studied.
Benson et al. \cite{benson2018simplicial} examined the interactions among a fixed number of nodes, with a focus on their relations with the emergence of a hyperedge containing all the nodes.
Lee et al. \cite{lee2020hypergraph} inspected the overlaps between three hyperedges, which they categorize into $26$ patterns called hypergraph motifs (h-motifs).
Comparing the relative counts of each h-motif's instances revealed that local structures are particularly similar between hypergraphs from the same domain but different across domains.
In h-motifs, however, temporal dynamics are completely ignored.

This line of research has also revealed that specialized analysis tools (e.g., h-motifs \cite{lee2020hypergraph} and multi-level decomposition \cite{do2020structural}) are useful for extracting unique high-order information that hypergraphs convey and also for coping with additional complexity due to the flexibility in the size of hyperedges.
Simply utilizing graph analysis tools (e.g., network motifs \cite{milo2002network}) after converting hypergraphs into pairwise graphs is often limited in addressing the above challenges \cite{lee2020hypergraph,yoon2020much}.

Motivated by interesting patterns that temporal network motifs revealed in ordinary graphs~\cite{paranjape2017motifs,li2018temporal,kovanen2011temporal,gurukar2015commit,redmond2013temporal}, 
we define $96$ \textit{temporal hypergraph motifs} (\shorts) for local pattern analysis of time-evolving hypergrpahs.
\shorts generalize the notion of static h-motifs, which completely ignore temporal information, and describe both relational and temporal dynamics among three connected hyperedges that arrive within a short time.
Specifically, given three connected hyperedges $e_i$, $e_j$, and $e_k$, all of which arrive within $\delta$ time units, \shorts describe their connectivity based on the emptiness of the seven subsets of them shown in Fig.~\ref{fig:example}(b).
In the temporal perspective, the relative arrival orders of $e_i$, $e_j$, and $e_k$ are taken into account, and thus patterns that are indistinguishable using static h-motifs can be characterized using \shorts. 

Given a temporal hypergraph, where a timestamp is attached to each hyperedge (see Fig.~\ref{fig:example}(a) for an example), we summarize its local structural and temporal characteristics using the relative occurrence of 96 \shorts' instances. 
That is, we obtain a vector of length $96$ regardless of the sizes and time spans of hypergraphs, and thus local characteristics of different hypergraphs can easily be compared. 

Another focus of this paper is the problem of counting \shorts' instances.
Since the number of three connected hyperedges can be orders of magnitude larger than the number of hyperedges, directly enumerating all of them is computationally prohibitive, especially for massive hypergraphs.
We develop \adv (\textbf{T}emporal \textbf{Hy}pergraph \textbf{M}otif C\textbf{e}nsus), which exactly counts each \short's instances while avoiding direct enumeration.
In our experiments, \adv is up to $\mathbf{2,163\times}$ \textbf{faster} than the direct extension of a recent exact temporal network motif counting algorithm \cite{paranjape2017motifs}, which enumerates every static h-motif in the induced static hypergraph.
\adv makes the best use of our two findings in real-world hypergraphs that temporal hyperedges tend to be (1) repetitive and (2) temporally local. These findings about duplicated (i.e., completely overlapped) hyperedges complement the findings in \cite{benson2018sequences}, which focus mainly on partial overlaps.


Using \shorts and \adv, we investigate $11$ real-world hypergraphs from $5$ distinct domains.
Our empirical study demonstrates that \shorts are informative, capturing both structural and temporal characteristics.
Specifically, using the counts of incident \shorts' instances as features brings up to $\mathbf{25.7\%}$ \textbf{improvement in the accuracy} of a hyperedge prediction task, compared to when static h-motifs are used instead of \shorts.
Moreover, \shorts reveal interesting patterns, including the striking similarity between hypergraphs from the same domain.


In summary, our contributions are as follow:
\begin{enumerate}
    \item \textbf{New concept:} We define $96$ temporal hypergraph motifs (\motifs) for characterizing local structures and dynamics in hypergraphs of various sizes.
    \item \textbf{Fast and exact algorithms:} We develop fast algorithms for exactly counting the instances of \motifs, and they are up to $2,163\times$ faster than baseline.
    \item \textbf{Empirical discoveries:} We demonstrate the usefulness of \motifs by uncovering the design principles of 11 real-world temporal hypergraphs from 5 different domains.
\end{enumerate}
\noindent \textbf{Reproducibility:} The source code and datasets used in this work are available at \url{https://github.com/geonlee0325/THyMe}.

In Section~\ref{sec:related} we review preliminaries and related prior works. 
In Section~\ref{sec:concept}, we present the concept of \shorts. 
In Section~\ref{sec:method}, we develop algorithms for counting the instances of \shorts. 
In Section~\ref{sec:experiments}, we empirically analyze real-world temporal hypergraphs through the lens of \shorts. 
Lastly, in Section~\ref{sec:conclusion}, we offer conclusions.

\begin{figure*}[t]
    \vspace{-2mm}
	\centering
	\includegraphics[width=0.995\textwidth]{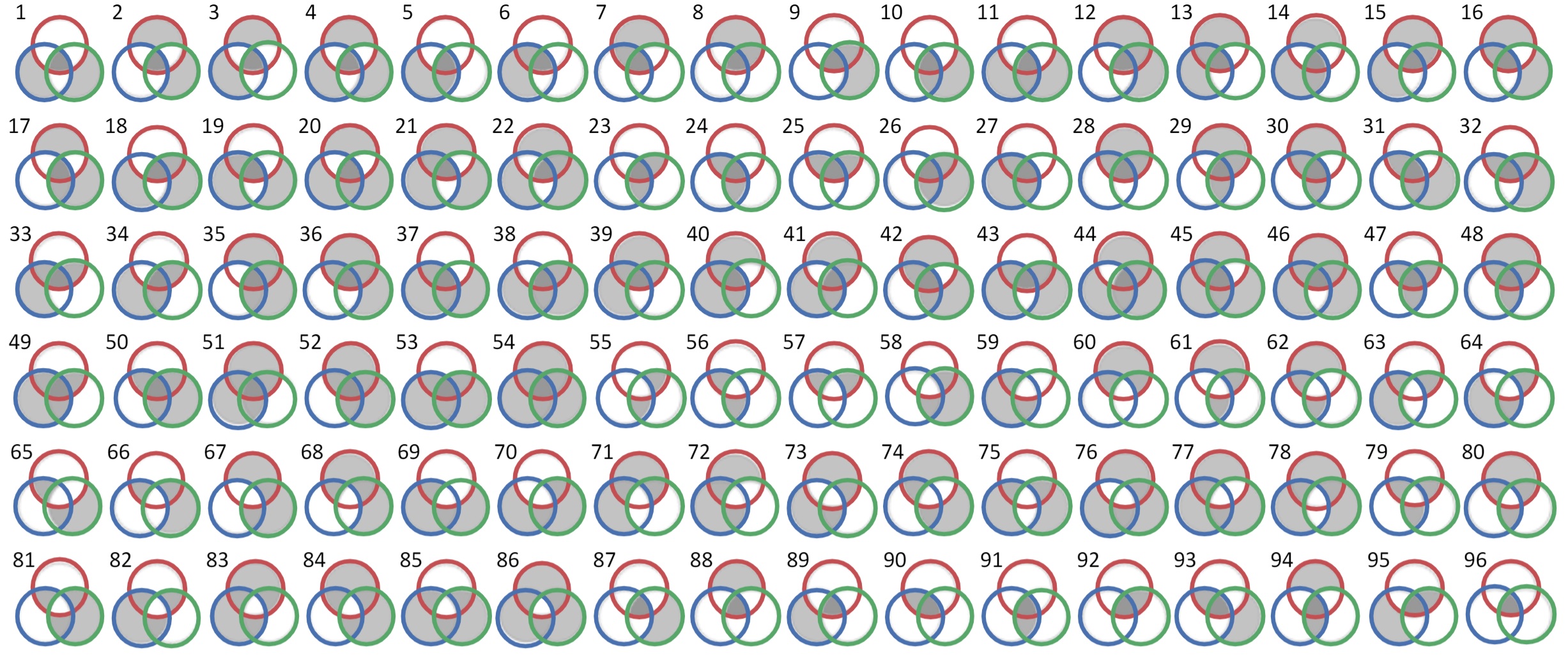}
	\caption{\label{fig:defn} \textbf{The 96 temporal hypergraph motifs (\shorts).} In each \short, the red hyperedge arrives first followed by the blue one and then the green one. Each of the $7$ distinct regions in the Venn diagram representation is colored white if it is empty, and it is colored grey if it is filled with at least one node. See Fig.~\ref{fig:example}(d) for an instance of \short 77.
	}
\end{figure*}
	
	\section{Preliminaries and Related Works}
	\label{sec:related}

In this section, we first review the concept of hypergraphs. Then, we introduce hypergraph motifs (h-motifs), which is designed for static hypergraphs.
Lastly, we discuss other related works.
Refer to Table~\ref{tab:notations} for the frequently-used notations.

\begin{table}[t!]
	\begin{center}
		\caption{\label{tab:notations}Frequently-used notations.}
		\scalebox{0.90}{
			\begin{tabular}{c|l}
				\toprule
				\textbf{Notation} & \textbf{Definition}\\
				\midrule
				$T=(V,\mathcal{E})$ & temporal hypergraph with temporal hyperedges $\mathcal{E}$\\
				$G_{T}=(V,E_{\mathcal{E}})$ & induced static hypergraph of the temporal hypergraph $T$\\
				\midrule
				$e_i=(\tilde{e}_i,t_i)$ & temporal hyperedge with nodes $\tilde{e}_i$ arrived at time $t_i$\\
				$I(\tilde{e})$ & set of temporal hyperedges whose nodes are $\tilde{e}$\\
				$h(\tilde{e}_i,\tilde{e}_j,\tilde{e}_k)$ & \motif corresponding to an instance $\langle e_i,e_j,e_k \rangle$\\
				\midrule
				$P=(V_P,E_P)$ & projected graph in \naive\\
				$Q=(V_Q,E_Q,t_Q)$ & projected graph in \adv\\
				\bottomrule %
			\end{tabular}}
	\end{center}
\end{table}

\subsection{Basic Concepts: Static and Temporal Hypergraphs}
A \textit{hypergraph} $G=(V,E)$ consists of a set of nodes $V=\{v_1,...,v_{|V|}\}$ and a set of hyperedges $E=\{\tilde{e}_1,...,\tilde{e}_{|E|}\}$. Each \textit{hyperedge} $\tilde{e}\in E$ is a non-empty set of an arbitrary number of nodes. 
A \textit{temporal hypergraph} $T=(V,\mathcal{E})$ on a node set $V$ is an ordered sequence of \textit{temporal hyperedges}. Each $i$\textsuperscript{th} temporal hyperedge $e_i=(\tilde{e}_i,t_i)$ where $\tilde{e}_i\subseteq V$ is the set of nodes and $t_i$ is the time of arrival.
Two distinct temporal hyperedges $e_i=(\tilde{e}_i,t_i)$ and $e_j=(\tilde{e}_j,t_j)$ are \textit{duplicated} if they share exactly same set of nodes, i.e., $\tilde{e}_i=\tilde{e}_j$.
We assume the sequence is ordered and timestamps are unique, i.e., if $i<j$, then $t_i<t_j$.
We denote the set of temporal hyperedges whose nodes are  $\tilde{e}$ (i.e., those \textit{inducing} $\tilde{e}$) by $I(\tilde{e}):=\{e_i=(\tilde{e}_i,t_i)\in \mathcal{E}: \tilde{e}_i =\tilde{e}\}$.
The temporal hypergraph $T$ induces a static hypergraph $G_T=(V,E_{\mathcal{E}})$ where timestamps and duplicated temporal hyperedges are ignored. 
That is, a hyperedge $\tilde{e}\in E_{\mathcal{E}}$ in $G_T$ exists if and only if $I(\tilde{e})\neq \emptyset$.
Notably, the number of temporal hyperedges is typically much larger than that of static hyperedges in the induced hypergraph, i.e., $|\mathcal{E}|\gg |E_{\mathcal{E}}|$.


\subsection{Static Hypergraph Motifs (h-motifs)}
\textit{Hypergraph motifs} (h-motifs)~\cite{lee2020hypergraph} are tools for understanding the local structural properties of static hypergraphs. 
Given three connected hyperedges, h-motifs describe their connectivity patterns by the emptiness of each of seven subsets: (1) $\tilde{e}_i \setminus \tilde{e}_j \setminus \tilde{e}_k$, (2) $\tilde{e}_j \setminus \tilde{e}_k \setminus \tilde{e}_i$, (3) $\tilde{e}_k \setminus \tilde{e}_i \setminus \tilde{e}_j$, (4) $\tilde{e}_i \cap \tilde{e}_j \setminus \tilde{e}_k$, (5) $\tilde{e}_j \cap \tilde{e}_k \setminus \tilde{e}_i$, (6) $\tilde{e}_k \cap \tilde{e}_i \setminus \tilde{e}_j$, and (7) $\tilde{e}_i \cap \tilde{e}_j \cap \tilde{e}_k$.
While there can exist $2^7$ possible cases of emptiness, $26$ cases of them are considered after excluding symmetric, duplicated, and disconnected ones.
Since non-pairwise interactions among the hyperedges (such as $\tilde{e}_i \cap \tilde{e}_j \cap \tilde{e}_k$) are taken into account, h-motifs effectively captures the high-order information of the overlapping patterns of the hyperedges.
It is shown empirically that their occurrences in the real-world hypergraphs are significantly different from those in randomized hypergraphs. Moreover, the relative occurrences are particularly similar between hypergraphs from the same domain, while they are distinct between hypergraphs from different domains.
Note that h-motifs, which is originally designed for static hypergraphs, completely ignore temporal information. 

	\subsection{Other Related Works}

In this subsection, we review prior works on network motifs and empirical analysis of hypergraphs. 

\smallsection{Network Motifs.}
Network motifs are fundamental building blocks of real-world graphs~\cite{shen2002network,milo2002network}.
Their relative occurrences in real-world graphs are significantly different from those in randomized ones~\cite{milo2002network} and unique within each domain~\cite{milo2004superfamilies}.
While they were originally defined on a static graph, they have been extended to 
temporal~\cite{paranjape2017motifs}, heterogeneous~\cite{rossi2020heterogeneous,li2018temporal}, and bipartite~\cite{borgatti1997network} graphs, as well as hypergraphs~\cite{lee2020hypergraph}. Their usefulness has been demonstrated in a wide range of graph applications including community detection~\cite{benson2016higher,li2019edmot,tsourakakis2017scalable,yin2017local,arenas2008motif}, ranking~\cite{zhao2018ranking}, and embedding~\cite{yu2019rum,rossi2018higher,rossi2020structural,lee2019graph,rossi2018deep}.

\smallsection{Temporal Network Motifs:}
The notion of network motifs has been extended to temporal networks to describe patterns in sequences of temporal edges. 
Several definitions of temporal motifs have been used, and most of them consider the \textit{temporal connectivity} between the edges. 
In \cite{kovanen2011temporal} and \cite{gurukar2015commit}, they consider $\delta$-adjacency between temporal edges. That is, every consecutive edges should share a node and arrive within in $\delta$ time units. 
Several counting algorithms for such patterns have been proposed~\cite{redmond2013temporal,kovanen2011temporal,gurukar2015commit}.
Another definition of temporal motifs 
describes patterns of sequences of temporal edges where all edges arrive within $\delta$ time units~\cite{paranjape2017motifs} while taking their relative arrival orders into consideration.
In this work, we define \shorts based on the notion of temporal motifs defined in \cite{paranjape2017motifs} due to its simplicity and effectiveness.

\smallsection{Empirical Analysis of Real-world Hypergraphs:}
Empirical analysis of global~\cite{lee2021hyperedges,do2020structural} and local~\cite{benson2018simplicial,lee2020hypergraph} structural patterns and temporal patterns~\cite{kook2020evolution,benson2018simplicial,benson2018sequences} of real-world hypergraphs has been performed, as discussed in detail in Section~\ref{sec:intro}. 
	

	\section{Proposed Concepts}
	\label{sec:concept}
	In this section, we propose \textit{temporal hypergraph motifs} (\motifs), which are tools for understanding the local structural and temporal characteristics of temporal hypergraphs. 
We introduce the definition and their relevant concepts.

\smallsection{Definition:}
\motifs describe structural and temporal patterns in sequences of three connected temporal hyperedges that are close in time. 
Note that three hyperedges are \textit{connected} if and only if one among them overlaps with the others.
Specifically, given three connected temporal hyperedges $\langle e_i=(\tilde{e}_i,t_i)$, $e_j=(\tilde{e}_j,t_j)$, $e_k=(\tilde{e}_k,t_k)\rangle$ where $t_i<t_j<t_k$ and $t_k-t_i \leq \delta$ (i.e., they arrive within a predefined time interval $\delta$), \motifs describe the emptiness of the $7$ subsets: (1) $\tilde{e}_i \setminus \tilde{e}_j \setminus \tilde{e}_k$, (2) $\tilde{e}_j \setminus \tilde{e}_k \setminus \tilde{e}_i$, (3) $\tilde{e}_k \setminus \tilde{e}_i \setminus \tilde{e}_j$, (4) $\tilde{e}_i \cap \tilde{e}_j \setminus \tilde{e}_k$, (5) $\tilde{e}_j \cap \tilde{e}_k \setminus \tilde{e}_i$, (6) $\tilde{e}_k \cap \tilde{e}_i \setminus \tilde{e}_j$, and (7) $\tilde{e}_i \cap \tilde{e}_j \cap \tilde{e}_k$.
That is, in the structural aspect, \motif 
describes the emptiness of the seven distinct regions in the Venn diagram representation (see Fig.~\ref{fig:example}(b)), effectively capturing the high-order connectivity among three hyperedges.
In the temporal aspects, \motifs take the relative arrival orders of three hyperedges and their time interval into consideration.
While there can exist $2^7$ possible cases of emptiness, we consider $96$ cases of them, which are called \motif 1 to \motif 96, after excluding those describing disconnected hyperedges.
We visualize the 96 \shorts in Fig.~\ref{fig:defn}.
Recall that static h-motifs completely ignore temporal information, and also assume that every hyperedge is unique, while \motifs also describe the patterns among duplicated temporal hyperedges.
Thus, while static h-motifs distinguish only $26$ different patterns, \motifs distinguish $96$ different patterns by considering temporal dynamics in addition to connectivity. 

\smallsection{Instance of \motifs:}
A sequence $\langle e_i,e_j,e_k \rangle$ of three temporal hyperedges is an \textit{instance} of \motif $t$ if their relational and temporal dynamics are described by \motif $t$ (see Fig.~\ref{fig:example}(d) for an example).
For each instance $\langle e_i,e_j,e_k \rangle$, we denote its corresponding \motif by $h(\tilde{e}_i,\tilde{e}_j,\tilde{e}_k)$.

\smallsection{Triple, Pair, and Single Inducing \motifs:}
The 96 \motifs can be categorized into three types based on the number of underlying static hyperedges. 
A \motif is \textit{\triple} if underlying hyperedges in its instance $\langle e_i,e_j,e_k \rangle$ are distinct (i.e., $\tilde{e}_i \neq \tilde{e}_j$, $\tilde{e}_j \neq \tilde{e}_k$, and $\tilde{e}_k \neq \tilde{e}_i$), as in \shorts $1$-$86$.
If two are duplicated while the remaining one is different, as in \shorts $87$-$95$, it is \textit{\pair}.
If all three hyperedges are duplicated, as in \short $96$, it is \textit{\single}.

    \section{Counting Algorithms}
    \label{sec:method}
    In this section, we describe methodologies for exactly counting the instances of each \motifs in the input temporal hypergraph.
We first present \wsdmshort, which extends a recent exact counting algorithm \cite{paranjape2017motifs} for temporal network motifs.
Then, we describe \naive, a preliminary version of our proposed algorithm \adv.
Lastly, we propose \adv (\textbf{T}emporal \textbf{Hy}pergraph \textbf{M}otif C\textbf{e}nsus), a fast and efficient algorithm that addresses the limitations of the previous ones. 

\smallsection{Remarks:}
The problem of counting \motifs has additional technical challenges while it bears some similarity with counting static h-motifs or temporal network motifs. 
First, the number of temporal hyperedges is typically much larger than that of hyperedges in the underlying static hypergraph. 
For example, the $11$ considered real-world temporal hypergraphs (see Section~\ref{sec:experiments:settings}) have up to $1.2-22.0\times$ more hyperedges than the underlying static ones.
This incurs significant bottlenecks of enumeration methods, and thus fast algorithms are demanded.
Temporal network motifs are defined only by pairwise interactions among a fixed number of nodes and their timestamps.
However, \motifs are defined not just by pairwise interactions but also by non-pairwise interactions among three hyperedges, in addition to their timestamps.


\subsection{\wsdm (\wsdmshort): Extension of \cite{paranjape2017motifs}}
We present \wsdm (\wsdmshort), which is a baseline approach for counting the instances of each \motif in the input temporal hypergraph $T$.

\smallsection{Counting \wsdmshort:}
Given an input temporal hypergraph $T=(V, \mathcal{E})$, \wsdmshort enumerates the instances of static h-motifs in the induced static hypergraph $G_T=(V,E_{\mathcal{E}})$.
This step can be processed by using an existing algorithm provided in \cite{lee2020hypergraph}.
For each instance $\{\tilde{e}_i,\tilde{e}_j,\tilde{e}_k\}$ of static h-motif in $G_T$, \wsdmshort counts the instances of each \motifs whose temporal hyperedges (a) induce the static h-motif instance and (b) arrive within $\delta$ time. 
To this end, we adapt the dynamic programming scheme provided by \cite{paranjape2017motifs}, as described in detail in Appendix A.


\begin{figure}[t]
	\centering
	\includegraphics[width=0.50\textwidth]{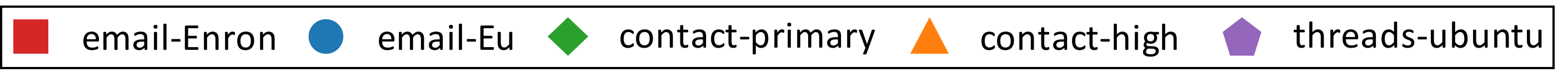}
	\begin{subfigure}[b]{.155\textwidth}
    	\includegraphics[width=0.99\linewidth]{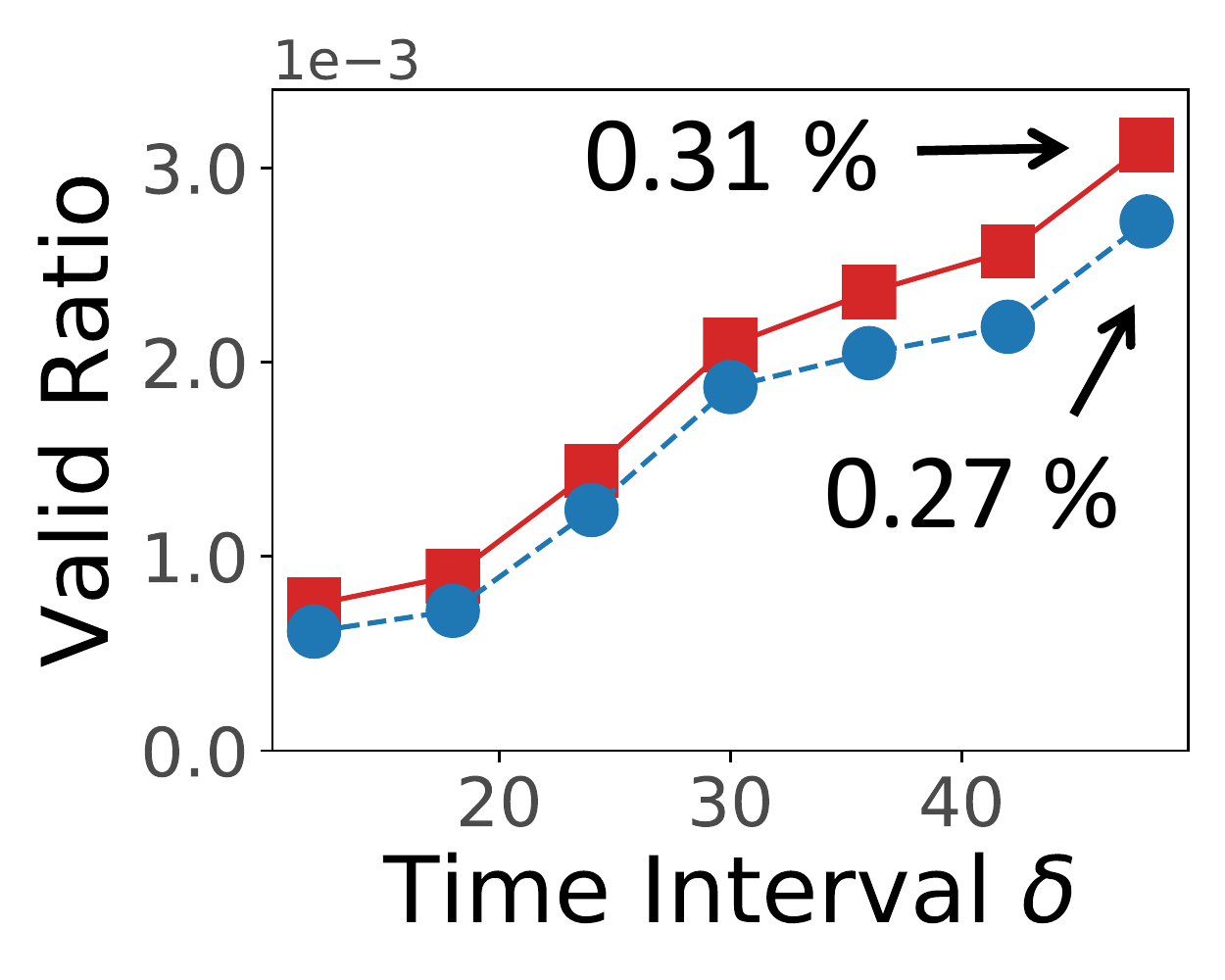}
    	\vspace{-15pt}
    	\caption{\small{\texttt{email}}}
	\end{subfigure}
	\begin{subfigure}[b]{.155\textwidth}
    	\includegraphics[width=0.99\linewidth]{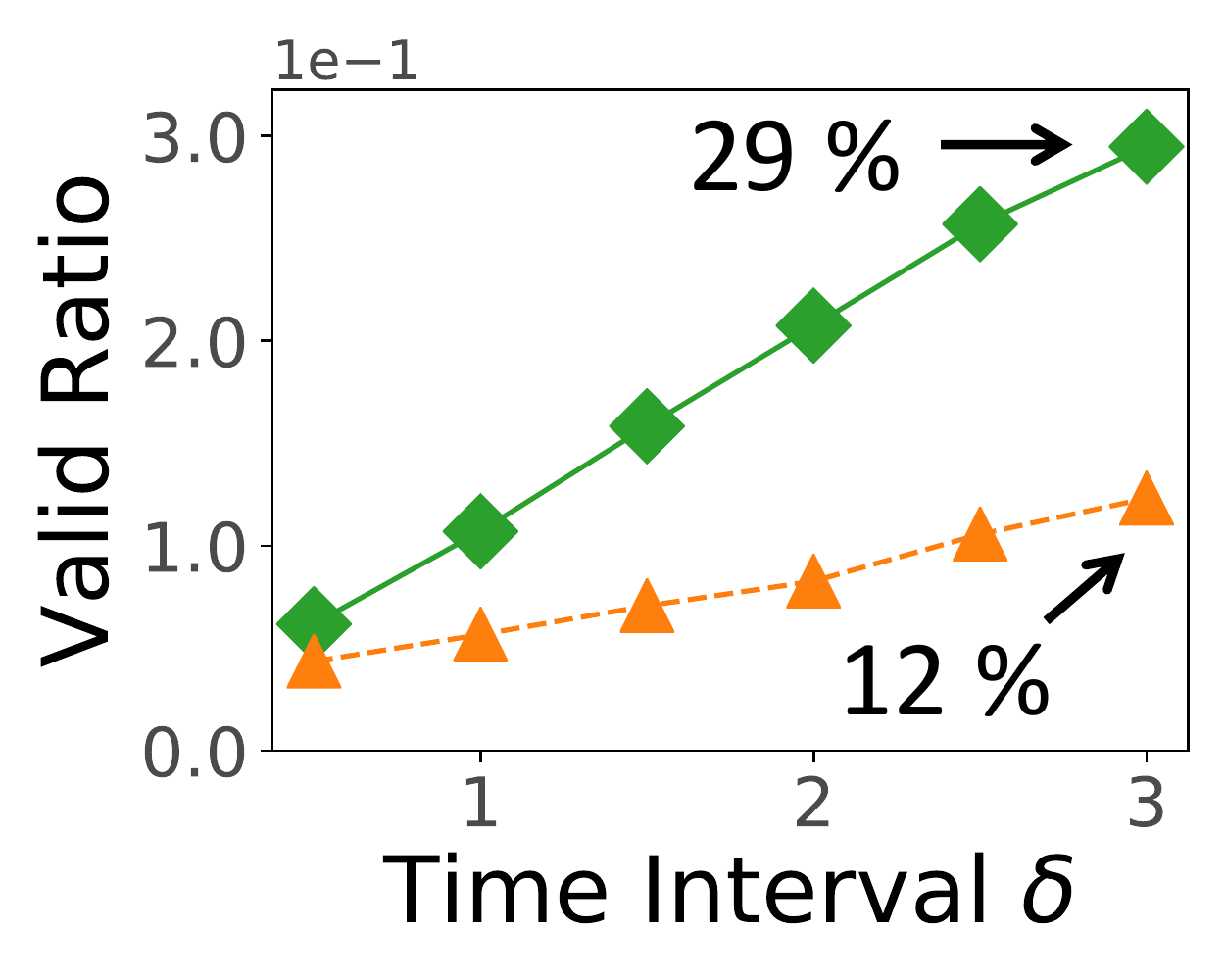}
    	\vspace{-15pt}
    	\caption{\small{\texttt{contact}}}
	\end{subfigure}
	\begin{subfigure}[b]{.155\textwidth}
    	\includegraphics[width=0.99\linewidth]{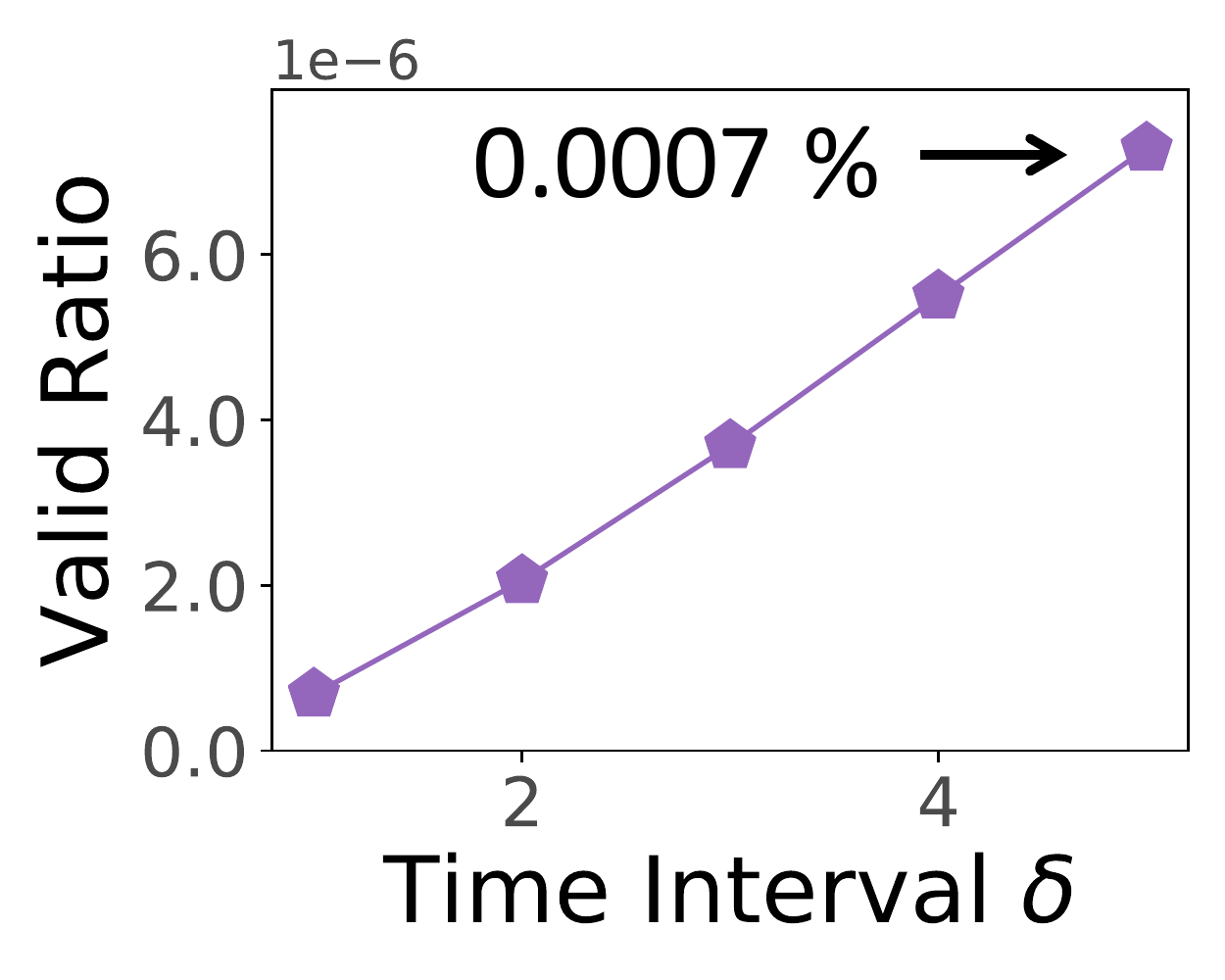}
    	\vspace{-15pt}
    	\caption{\small{\texttt{threads}}}
	\end{subfigure}
	\caption{Only a small fraction of static h-motifs' instances in the induced static hypergraphs are induced by any valid instance of \motifs. 
	Results in small datasets where the instances of static h-motifs can be exactly counted are reported.
	\label{fig:valid}}
\end{figure}

\smallsection{Limitations of \wsdmshort:}
Using dynamic programming, \wsdmshort avoids enumerating over all instances of \motifs. 
However, it still enumerates all instances of static h-motifs in the induced hypergraph $G_T$, most of which however are not induced by any valid instance of \shorts, as seen in Fig.~\ref{fig:valid}.
For example, in \texttt{threads-ubuntu}, only $0.0007\%$ of the static h-motifs instances are induced by any valid instance of \motifs when $\delta$ is $5$ hours.
That is, \wsdmshort enumerates every three connected hyperedges in $G_T$, ignoring any temporal information, while we are interested only in three connected temporal hyperedges that arrive within in a short period of time.

\begin{algorithm}[t]
    \small
	\caption{\naive: 
	Preliminary Algorithm 
	\label{alg:naive}}
	\DontPrintSemicolon
	\SetKwInOut{Input}{Input}
    \SetKwInOut{Output}{Output}
    \SetKwFunction{Finsert}{insert}
    \SetKwFunction{Fremove}{remove}
    \SetKwComment{Comment}{$\triangleright$}{}
    \Input{(1) temporal hypergraph: $T=(V,\mathcal{E})$\\(2) time interval $\delta$}
    \Output{\# of each temporal h-motif $t$'s instances: $M[t]$}
    \vspace{2pt}
    
    $M \leftarrow$ map initialized to 0\\
    $P = (V_P=\varnothing,E_P=\varnothing)$\label{alg:naive:initP}\\
    $w_s \leftarrow 1$\\
    \vspace{2pt}
    
    \For{\upshape\textbf{each} temporal hyperedge $e_i=(\tilde{e}_i,t_i)\in\mathcal{E}$}{
        \Finsert{$e_i$}~\label{alg:naive:insert}\\
        \While{$t_{w_s} + \delta < t_i$}{
            \Fremove{$e_{w_s}$}~\label{alg:naive:remove}\\
            $w_s \leftarrow w_s + 1$~\label{alg:naive:remove2}\\
        }
        $S \leftarrow$ \hspace{-2mm} set of $3$ connected temporal hyperedges including $e_i$\label{alg:naive:enum} \\ 
        \For{\upshape\textbf{each} instance $\langle e_j,e_k,e_i \rangle\in S$}{
            $M[h(\tilde{e}_j,\tilde{e}_k,\tilde{e}_i)]$ += 1\label{alg:naive:inc}
        }
    }
    \Return{M}\\
    \vspace{3pt}

    \SetKwProg{myproc}{Procedure}{}{}
    \myproc{\Finsert{$e_i=(\tilde{e}_i,t_i)$}}{
        $V_P \leftarrow V_P \cup \{e_i\}$\label{alg:naive:insert1}\\
        $N_{e_i} \leftarrow \{e : e \in V_P \setminus \{e_i\} \ \text{and} \ \tilde{e}_i \cap \tilde{e}\neq\varnothing\}$\label{alg:naive:insert2}\\
        $E_P \leftarrow E_P \cup \{(e_i,e) : e \in N_{e_i}\}$\label{alg:naive:insert3}
    }
    \vspace{1pt}
    \myproc{\Fremove{$e_i=(\tilde{e}_i,t_i)$}}{
        $V_P \leftarrow V_P \setminus \{e_i\}$\label{alg:naive:remove1}\\
        $N_{e_i} \leftarrow \{e : e \in V_P \ \text{and} \ \tilde{e}_i \cap \tilde{e} \neq \varnothing \}$\label{alg:naive:remove2}\\
        $E_P \leftarrow E_P \setminus \{(e_i,e) : e \in N_{e_i}\}$\label{alg:naive:remove3}\\
    }
\end{algorithm}

\subsection{\naive: Preliminary Version of the Proposed Algorithm} 
To address the limitations of \wsdmshort, we present \naive, a preliminary version of our proposed algorithm \adv.
\naive directly enumerates each instance of \motifs, instead of those of static h-motifs, to avoid unnecessary search.
To this end, \naive concisely considers the temporal hyperedges that occur in the $\delta$-sized temporal window.
In response to the arrival of a new temporal hyperedge $e_i$ at time $t_i$, the temporal window moves to $[t_i-\delta, t_i]$.
It maintains only a succinct projected graph $P=(V_P,E_P)$ that represents the connectivity between the temporal hyperedges that occur within the current temporal window.
As the window moves, the projected graph $P$ is incrementally updated, reflecting the changes of the current temporal hyperedges.
Using $P$, \naive exhaustively enumerates the instances of \motifs.

\smallsection{Projected Graph in \naive:}
The projected graph $P=(V_P,E_P)$ is a graph where each node is a temporal hyperedge and two nodes are connected as an edge if their corresponding temporal hyperedges share any nodes.
In \naive, $P$ is maintained on the fly, with response to the temporal hyperedges that either enter or exit the sliding time window. The update schemes are described as \texttt{insert} and \texttt{remove}, respectively, in Algorithm~\ref{alg:naive}.
In \texttt{insert}, a temporal hyperedge $e_i$ is added as a node (line~\ref{alg:naive:insert1}) and its neighbors (i.e., those in $V_P$ that are adjacent to $e_i$) are joined by edges (lines~\ref{alg:naive:insert2}-\ref{alg:naive:insert3}).
In \texttt{remove}, a temporal hyperedge $e_i$, as well as its incident edges are removed from $V_P$ and $E_P$, respectively (lines~\ref{alg:naive:remove1}-\ref{alg:naive:remove3}).

\smallsection{Counting in \naive:}
The counting procedure of \naive is described in Algorithm~\ref{alg:naive}.
The sets of nodes and edges of the projected graph $P$ are initialized to empty maps, i.e., $V_P=\varnothing$ and $E_P=\varnothing$ (line~\ref{alg:naive:initP}).
Once a temporal hyperedge $e_i=(\tilde{e}_i,t_i)\in\mathcal{E}$ arrives, the temporal window is moved to $[t_i-\delta,t_i]$ and the projected graph $P$ is updated accordingly, as described above. 
Then, it enumerates the instances of three connected nodes in $P$, which corresponds to the instances of \motifs of $T$ containing $e_i$ (line~\ref{alg:naive:enum}). 
For each instance $\langle e_j,e_k,e_i \rangle$ of \motif $t$, the corresponding count $M[t]$ is incremented (line~\ref{alg:naive:inc}). 

\smallsection{Limitations of \naive:}
Though \naive avoids redundant search in the induced static hypergraph $G_T$, it directly enumerates every instance of \motifs in $T$.
Since the size of the temporal hypergraph is much larger than that of induced static hypergraph, counting the instances in temporal hypergraph can be more computationally challenging, especially when time interval $\delta$ is large.
Each temporal hyperedge within the temporal window corresponds to a unique node in the projected graph $P$ even when many temporal hyperedges are highly duplicated, as in real-world hypergraphs (see Section~\ref{sec:experiments:analysis}).


\begin{algorithm}[t]
    \small
	\caption{\adv: Proposed Algorithm
	\label{alg:adv}}
	\DontPrintSemicolon
	\SetKwInOut{Input}{Input}
    \SetKwInOut{Output}{Output}
    \SetKwFunction{Finsert}{insert}
    \SetKwFunction{Fremove}{remove}
    \SetKwFunction{FcombThree}{comb3}
    \SetKwFunction{FcombTwo}{comb2}
    \SetKwFunction{FcombOne}{comb1}
    \SetKwComment{Comment}{$\triangleright$}{}
    \Input{\normalsize{(1) temporal hypergraph: $T=(V,\mathcal{E})$\\(2) time interval $\delta$}}
    \Output{\normalsize{\# of each temporal h-motif $t$'s instances: $M[t]$}}
    \vspace{2pt}
    
    $M \leftarrow$ map initialized to 0\\
    $Q = (V_Q=\varnothing,E_Q=\varnothing,t_Q=\varnothing)$\label{alg:adv:Qinit}\\
    $w_s \leftarrow 1$\\
    \vspace{2pt}
    
    \For{\upshape\textbf{each} temporal hyperedge $e_i=(\tilde{e}_i,t_i)\in\mathcal{E}$}{
        \Finsert{$e_i$}\label{alg:adv:main:insert}\\
        \While{$t_{w_s} + \delta < t_i$}{
            \Fremove{$e_{w_s}$}\label{alg:adv:main:remove1}\\
            $w_s \leftarrow w_s + 1$\label{alg:adv:main:remove2}\\
        }
        $S \leftarrow$ set of $3$ connected static hyperedges including $\tilde{e}_i$\label{alg:adv:main:enum}\\ 
        \For{\upshape\textbf{each} instance $\{\tilde{e}_i,\tilde{e}_j,\tilde{e}_k\}\in S$}{
            \FcombThree{$\tilde{e}_i, \tilde{e}_j, \tilde{e}_k$}\label{alg:adv:main:comb3}\\
        }
        \For{\upshape\textbf{each} pair $(\tilde{e}_i,\tilde{e}_j)\in N_{\tilde{e}_i}$\label{alg:adv:main:pair}}{
            \FcombTwo{$\tilde{e}_i, \tilde{e}_j$}\label{alg:adv:main:comb2}
        }
        \FcombOne{$\tilde{e}_i$}\label{alg:adv:main:comb1}
    }
    \Return{M}\\
    \vspace{4pt}
    
    \SetKwProg{myproc}{Procedure}{}{}
    \myproc{\Finsert{$e_i=(\tilde{e}_i,t_i)$}}{
        \If{$\tilde{e}_i \notin V_Q$}{
            $V_Q \leftarrow V_Q \cup \{\tilde{e}_i\}$\label{alg:adv:insert1}\\
            $N_{\tilde{e}_i} \leftarrow \{\tilde{e} : \tilde{e}\in V_Q \setminus \{\tilde{e}_i\} \ \text{and} \ \tilde{e}_i \cap \tilde{e} \neq \varnothing\}$\label{alg:adv:insert2}\\
            $E_Q \leftarrow E_Q \cup \{(\tilde{e}_i,\tilde{e}) : \tilde{e} \in N_{\tilde{e}_i}\}$\label{alg:adv:insert3}\\
        }
        $t_Q(\tilde{e}_i) \leftarrow t(\tilde{e}_i) \cup \{t_i\}$\label{alg:adv:insert_time}\\
    }
    \vspace{2pt}
    \myproc{\Fremove{$e_i=(\tilde{e}_i,t_i)$}}{
        $t(\tilde{e}_i) \leftarrow t(\tilde{e}_i) \setminus \{t_i\}$\label{alg:adv:remove_time}\\
        \If{$t_Q(\tilde{e}_i)=\varnothing$}{
            $V_Q \leftarrow V_Q \setminus \{\tilde{e}_i\}$\label{alg:adv:remove1}\\
            $N_{\tilde{e}_i} \leftarrow \{\tilde{e} : \tilde{e}\in V_Q \ \text{and} \ e_i \cap e \neq \varnothing\}$\label{alg:adv:remove2}\\
            $E_Q \leftarrow E_Q \setminus \{(\tilde{e}_i,\tilde{e}) : \tilde{e} \in N_{\tilde{e}_i}\}$\label{alg:adv:remove3}\\
        }
    }
    \vspace{2pt}
    \myproc{\FcombThree{$\tilde{e}_i, \tilde{e}_j, \tilde{e}_k$}\label{alg:adv:comb3}}{
        $M[h(\tilde{e}_j, \tilde{e}_k, \tilde{e}_i)]$ += $\sum_{t\in t_Q(\tilde{e}_j), t'\in t_Q(\tilde{e}_k)} \mathds{1}[t<t']$\label{alg:adv:comb3_1}\\
        $M[h(\tilde{e}_k, \tilde{e}_j, \tilde{e}_i)]$ += $\sum_{t\in t_Q(\tilde{e}_j), t'\in t_Q(\tilde{e}_k)} \mathds{1}[t'<t]$\label{alg:adv:comb3_2}\\
    }
    \vspace{2pt}
    \myproc{\FcombTwo{$\tilde{e}_i, \tilde{e}_j$}}{
        $M[h(\tilde{e}_i, \tilde{e}_j, \tilde{e}_i)]$ += $\sum_{t\in t_Q(\tilde{e}_i) \setminus \{t_i\}, t'\in t_Q(\tilde{e}_j)} \mathds{1}[t<t']$\label{alg:adv:comb2_1}\\
        $M[h(\tilde{e}_j, \tilde{e}_i, \tilde{e}_i)]$ += $\sum_{t\in t_Q(\tilde{e}_i) \setminus \{t_i\}, t'\in t_Q(\tilde{e}_j)} \mathds{1}[t'<t]$\label{alg:adv:comb2_2}\\
        $M[h(\tilde{e}_j, \tilde{e}_j, \tilde{e}_i)]$ += $|t_Q(\tilde{e}_j)| \choose 2$\label{alg:adv:comb2_3}
    }
    \vspace{2pt}
    \myproc{\FcombOne{$\tilde{e}_i$}}{
        $M[h(\tilde{e}_i,\tilde{e}_i,\tilde{e}_i)]$ += $|t_Q(\tilde{e}_i)-\{t_i\}| \choose 2$\label{alg:adv:comb1}\\
    }
\end{algorithm}

\subsection{\adv: Advanced Version of the Proposed Algorithm} 
We present \adv, our proposed algorithm for exactly counting the instances of \motifs. 
\adv is faster and more efficient than \wsdmshort and \naive, as shown empirically in Section~\ref{sec:experiments},
by addressing their limitations as follows.

\begin{itemize}[leftmargin=*]
    \item \wsdmshort enumerates all instances of static h-motifs in the induced hypergraph $G_T$, where most of them are redundant, not induced by any instance of \motifs of the temporal hypergraph $T$. \adv selectively enumerates the h-motif instances and thus reduces the redundancy.
    \item \naive exhaustively enumerates all instances of \motifs. \adv reduces the enumeration by introducing an effective counting scheme.
    \item The projected graph $P$ maintained by \naive can be large since each temporal hyperedge is represented as a unique node. \adv maintains a projected graph $Q$ that is typically smaller than $P$. In $Q$, the same node can be shared by multiple temporal hyperedges. The motivation behind $Q$ is empirically demonstrated in Section~\ref{sec:experiments:analysis}.
\end{itemize}


\smallsection{Projected Graph in \adv:}
\adv maintains a projected graph $Q=(V_Q,E_Q,t_Q)$ composed of a set of nodes $V_Q$, a set of edges $E_Q$, and a map $t_Q$. 
Each node and edge represent a static hyperedge and a pair of static hyperedges that share any nodes, respectively. In addition, $t_Q$ maps a set of timestamps of temporal hyperedges inducing a particular static hyperedge. 
Notably, while each node in the projected graph $P$ used in \naive is a unique temporal hyperedge, $Q$ represents the connectivity between hyperedges in the induced static hypergraph $G_T$.
That is, duplicated temporal hyperedges can share the same node in $Q$, and thus the size of the graph can be much smaller than $P$, i.e., $|E_Q|<|E_P|$.

The update schemes of $Q$, \texttt{insert} and \texttt{remove} in Algorithm~\ref{alg:adv} add or delete nodes and their adjacent edges, respectively. 
More specifically, in \texttt{insert}, given a new temporal hyperedge $e_i=(\tilde{e}_i,t_i)$, its set of nodes $\tilde{e}_i$ is inserted as a new node, only if there do not exist any temporal hyperedges in the current temporal window whose nodes are $\tilde{e}_i$ (line~\ref{alg:adv:insert1}). Once the new node is inserted, their incident edges are created as well (lines~\ref{alg:adv:insert2}-\ref{alg:adv:insert3}). Finally, the timestamp $t_i$ is added in $t_Q(\tilde{e}_i)$ (line~\ref{alg:adv:insert_time}).
In \texttt{remove}, given a temporal hyperedge $e_i$ to be removed, it first deletes its timestamp $t_i$ from $t_Q(\tilde{e}_i)$ (line~\ref{alg:adv:remove_time}).
If the $e_i$ is the only temporal hyperedge in the current window whose node set is $\tilde{e}_i$, then $\tilde{e}_i$ and its incident edges are removed from $V_Q$ and $E_Q$, respectively (lines~\ref{alg:adv:remove1}-\ref{alg:adv:remove3}).

\smallsection{Counting in \adv:}
The counting procedure of \adv is described in Algorithm~\ref{alg:adv}. 
The sets of nodes and edges of the projected graph $Q$ are initialized to empty maps, i.e., $V_Q=\varnothing$ and $E_Q=\varnothing$ (line~\ref{alg:adv:Qinit}).
For each temporal hyperedge $e_i=(\tilde{e}_i,t_i)$, it moves the temporal window to $[t_i-\delta,t_i]$ and accordingly as described above. 
Once $Q$ is updated, \adv counts the instances of \motifs that contains $e_i$ and the previous temporal hyperedges. 
To minimize enumerations, \adv adapts effective counting schemes, \texttt{comb3}, \texttt{comb2}, and \texttt{comb1}, which compute the number of instances of \triple, \pair, and \single \motifs, respectively, as follows:

\begin{itemize}[leftmargin=*]
    \item \textbf{\Triple \motifs (lines~\ref{alg:adv:main:enum}-\ref{alg:adv:main:comb3}):} \adv first enumerates the instances of three connected hyperedges in $Q$ such that contains $\tilde{e}_i$ (line~\ref{alg:adv:main:enum}). For each set $\{\tilde{e}_i,\tilde{e}_j,\tilde{e}_k\}$ of three connected hyperedges, the number of instances of \motifs that contains $e_i$ is counted by timestamp combinations using \texttt{comb3} method. That is, since $e_i$ is the latest temporal hyperedge, the set $\{\tilde{e}_i,\tilde{e}_j,\tilde{e}_k\}$ can be induced by sequences of either $\langle e_x,e_y,e_i \rangle$ or $\langle e_y,e_x,e_i \rangle$ where $\tilde{e}_x=\tilde{e}_j$ and  $\tilde{e}_y=\tilde{e}_k$. 
    Since $t_x\in t_Q(\tilde{e}_j)$ and $t_y\in t_Q(\tilde{e}_k)$, the number of such instances can be computed by the number of timestamp combinations of $t_Q(\tilde{e}_j)$ and $t_Q(\tilde{e}_k)$ (lines~\ref{alg:adv:comb3_1}-\ref{alg:adv:comb3_2}).
    \item \textbf{\Pair \motifs (lines~\ref{alg:adv:main:pair}-\ref{alg:adv:main:comb2}):} \adv enumerates each edge $(\tilde{e}_i, \tilde{e}_j)$ in $Q$ that are adjacent to $\tilde{e}_i$, which can be induced by three different orders of sequences, $\langle e_x,e_y,e_i \rangle$, $\langle e_y,e_x,e_i \rangle$, and $\langle e_y,e_y,e_i \rangle$ where $\tilde{e}_x=\tilde{e}_i$ and $\tilde{e}_y=\tilde{e}_j$. Since $t_x\in t_Q(\tilde{e}_i) \setminus \{t_i\}$ and $t_y \in t_Q(\tilde{e}_j)$, the number of the sequences can be computed by the number of combinations of the set of these timestamps (lines~\ref{alg:adv:comb2_1}-\ref{alg:adv:comb2_2}).
    \item \textbf{\Single \motifs (line~\ref{alg:adv:main:comb1}):} \Single \motif, which consists of three duplicated temporal hyperedges, can be immediately counted using \texttt{comb1}. That is, a sequence $\langle e_x,e_y,e_i \rangle$ where $\tilde{e}_x=\tilde{e}_i$ and $\tilde{e}_y=\tilde{e}_i$ can be an instance of \single \motif. Since $t_x\in t_Q(\tilde{e}_i)\setminus \{t_i\}$, $t_y\in t_Q(\tilde{e}_i)\setminus \{t_i\}$, and $t_x<t_y$, the number of such instances is computed immediately (line~\ref{alg:adv:comb1}).
\end{itemize}

In Section~\ref{sec:experiments:analysis}, we share empirical observations supporting the intuition behind \adv. In addition, we provide the complexity analysis of \adv in the supplementary document~\cite{online2021appendix}.

    
    \section{Empirical Studies}
    \label{sec:experiments}
    
\begin{figure*}[t]
	\vspace{-2mm}
	\centering
	\includegraphics[width=0.4\textwidth]{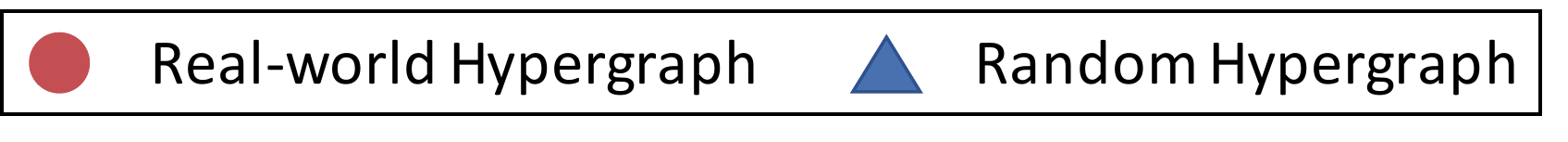}\\
	\begin{subfigure}[b]{.245\textwidth}
    	\includegraphics[width=0.99\linewidth]{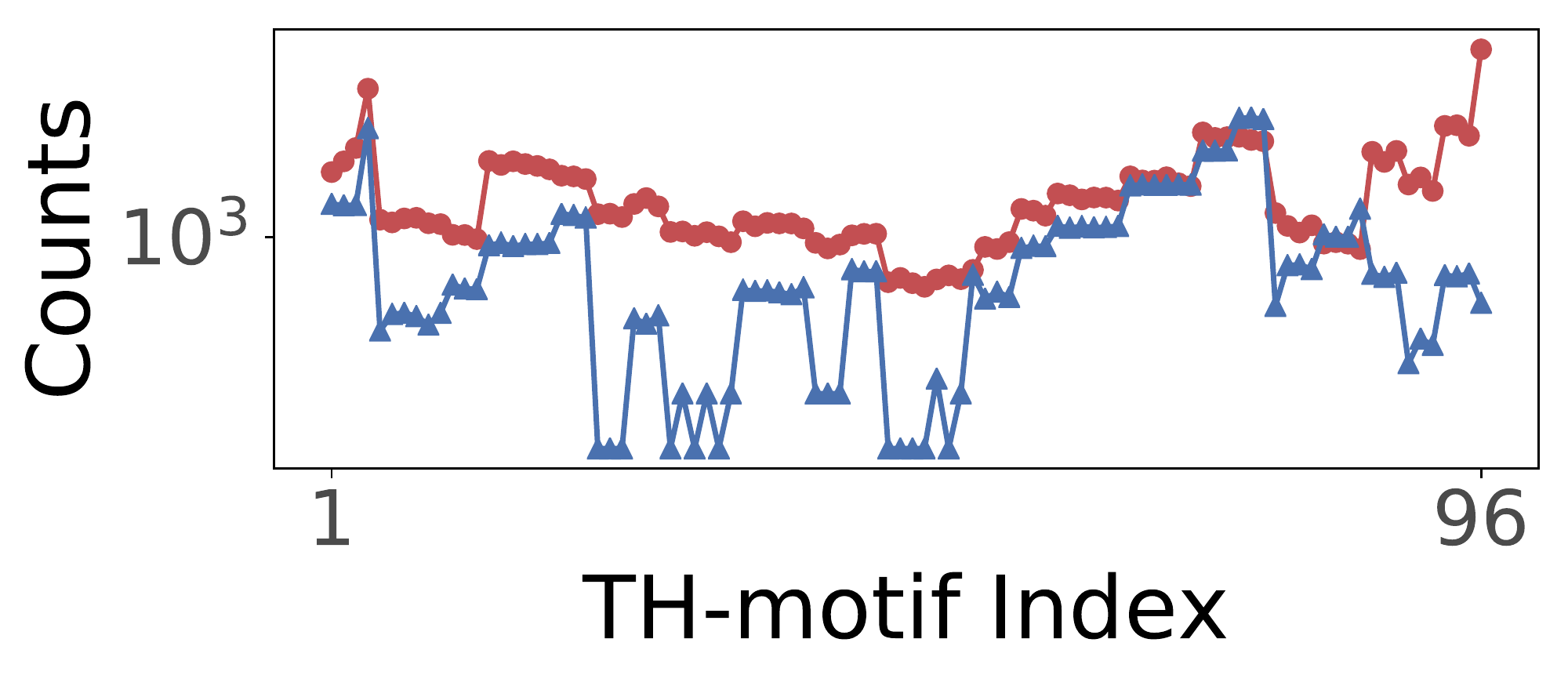}
    	\vspace{-14pt}
    	\caption{\texttt{email-Eu}}
	\end{subfigure}
	\hfill
	\begin{subfigure}[b]{.245\textwidth}
    	\includegraphics[width=0.99\linewidth]{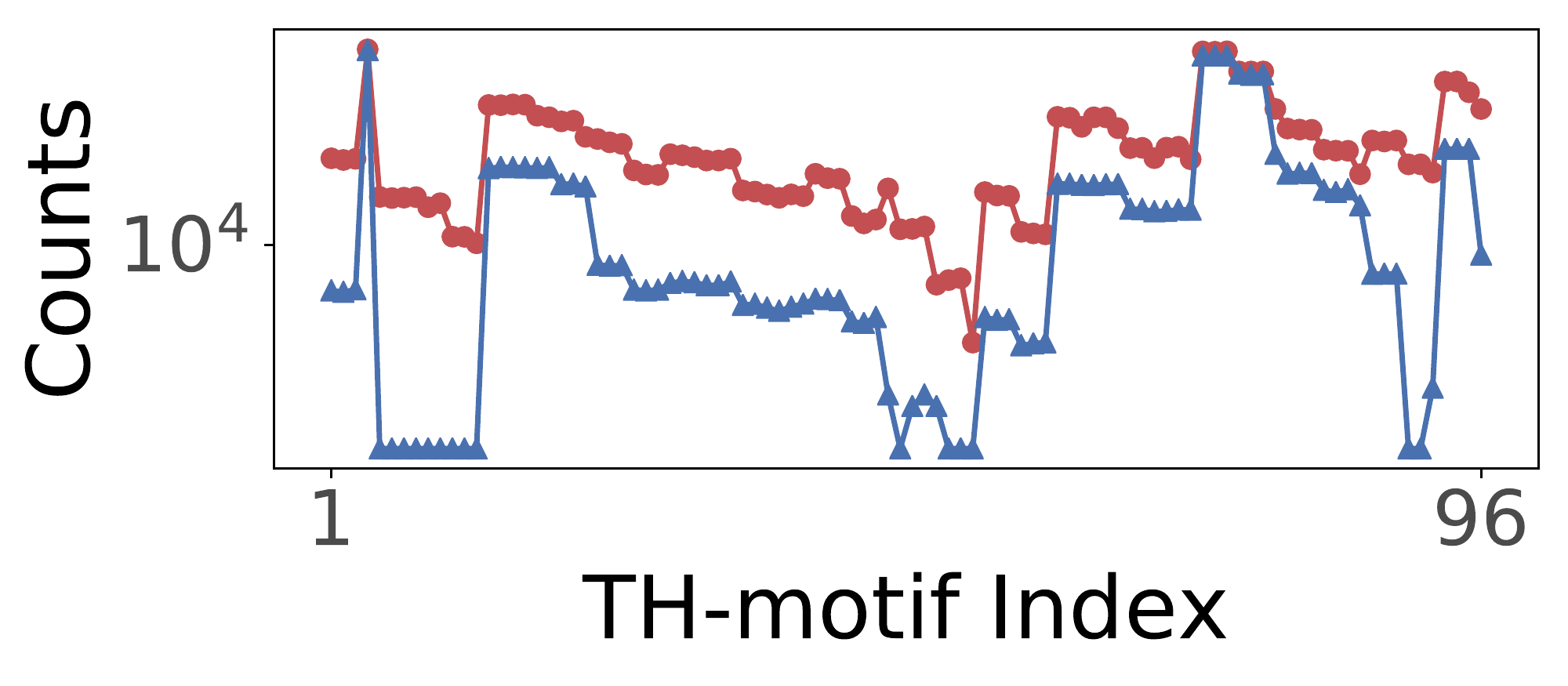}
    	\vspace{-14pt}
    	\caption{\texttt{contact-primary}}
	\end{subfigure}
	\hfill
	\begin{subfigure}[b]{.245\textwidth}
    	\includegraphics[width=0.99\linewidth]{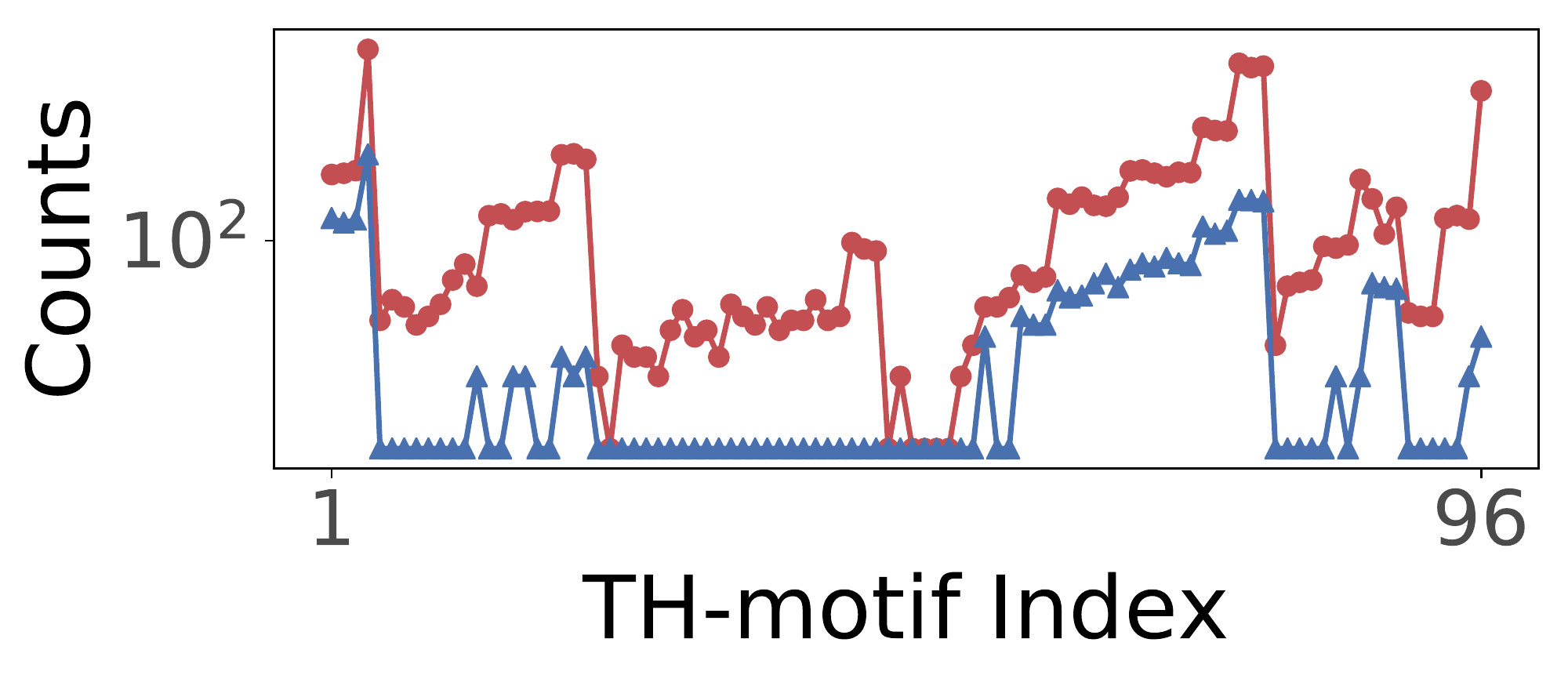}
    	\vspace{-14pt}
    	\caption{\texttt{threads-math}}
	\end{subfigure}
	\hfill
	\begin{subfigure}[b]{.245\textwidth}
    	\includegraphics[width=0.99\linewidth]{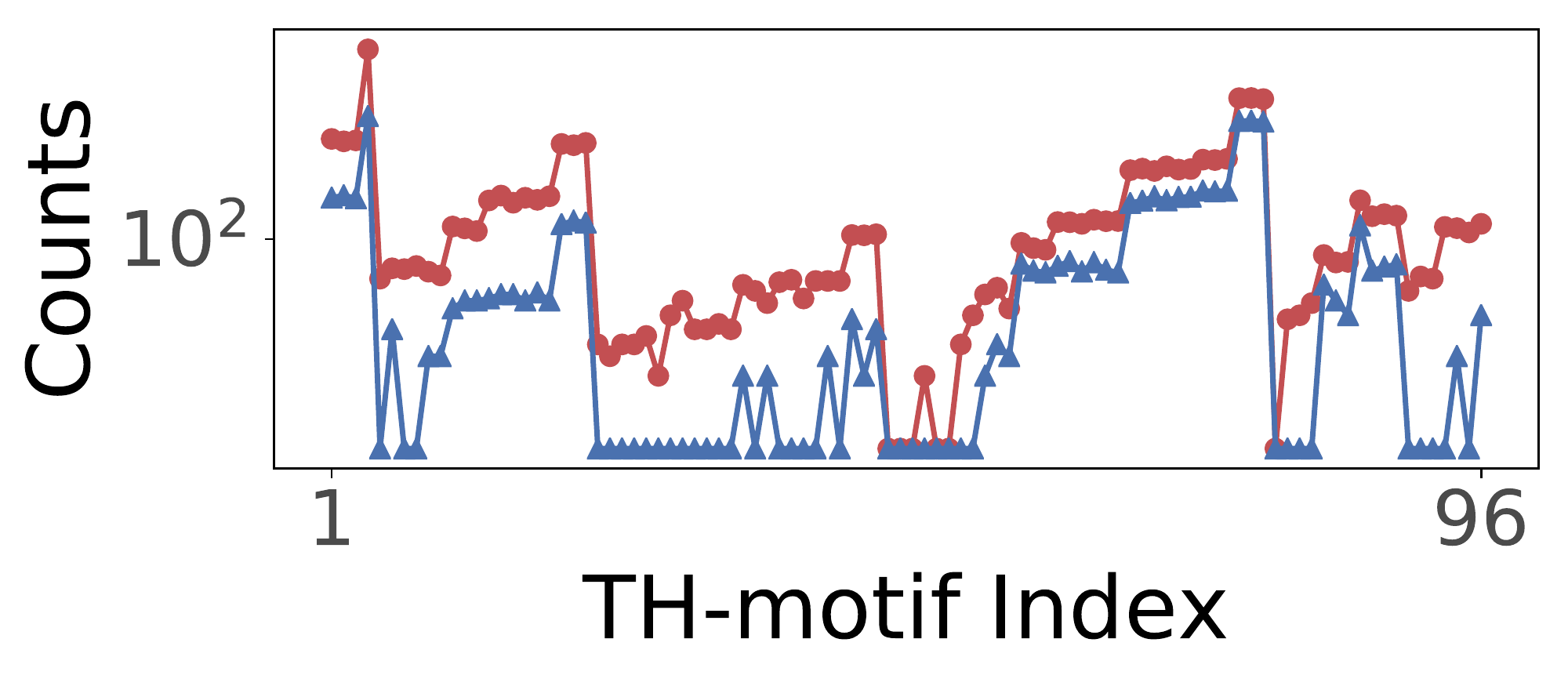}
    	\vspace{-14pt}
    	\caption{\texttt{tags-ubuntu}}
	\end{subfigure}
	\caption{The distribution of the number of \motifs' instances in real-world temporal hypergraphs and that in randomized temporal hypergraphs are significantly different. We set $\delta$ to 1 hour. We do not directly compare the distributions from the coauthorship datasets since their timestamp units are years. We provide the distributions in \cite{online2021appendix}. \label{fig:counts}}
\end{figure*}

\begin{figure}[t]
	\vspace{-2mm}
	\centering
	\includegraphics[height=0.55cm]{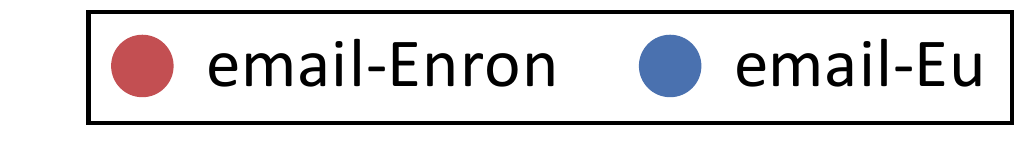}\\
	\vspace{-2.5pt}
	\includegraphics[width=0.95\linewidth]{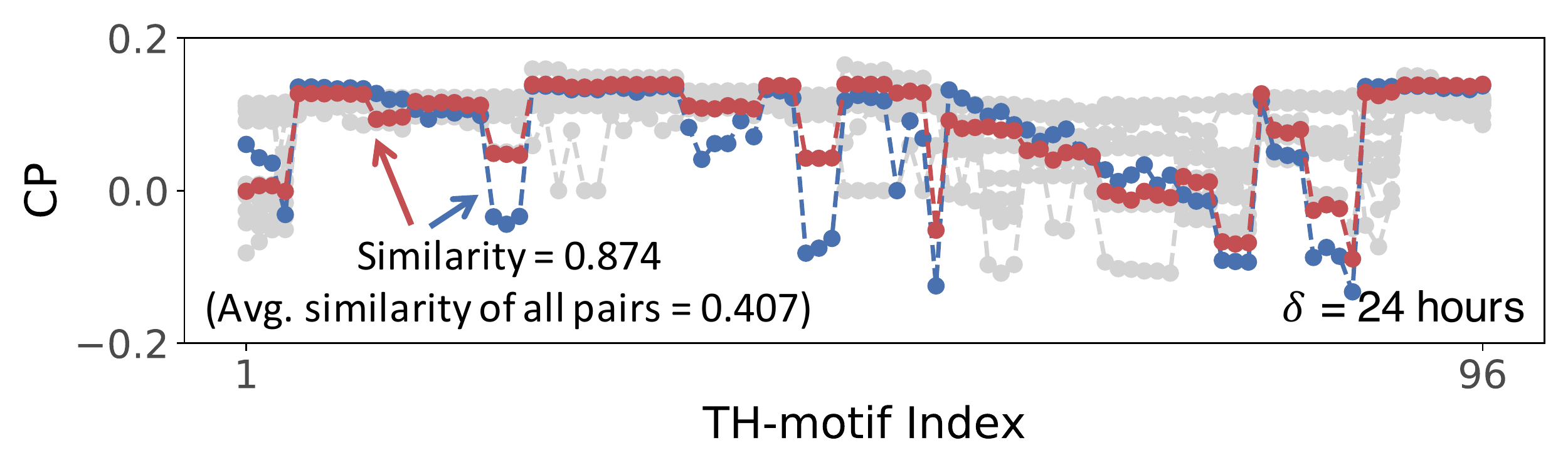}\\
	\vspace{5pt}
	\includegraphics[height=0.55cm]{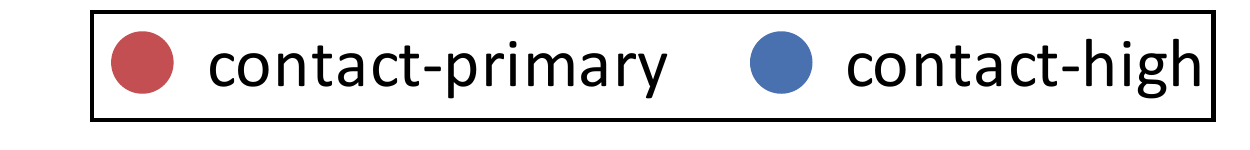}\\
	\vspace{-2.5pt}
	\includegraphics[width=0.95\linewidth]{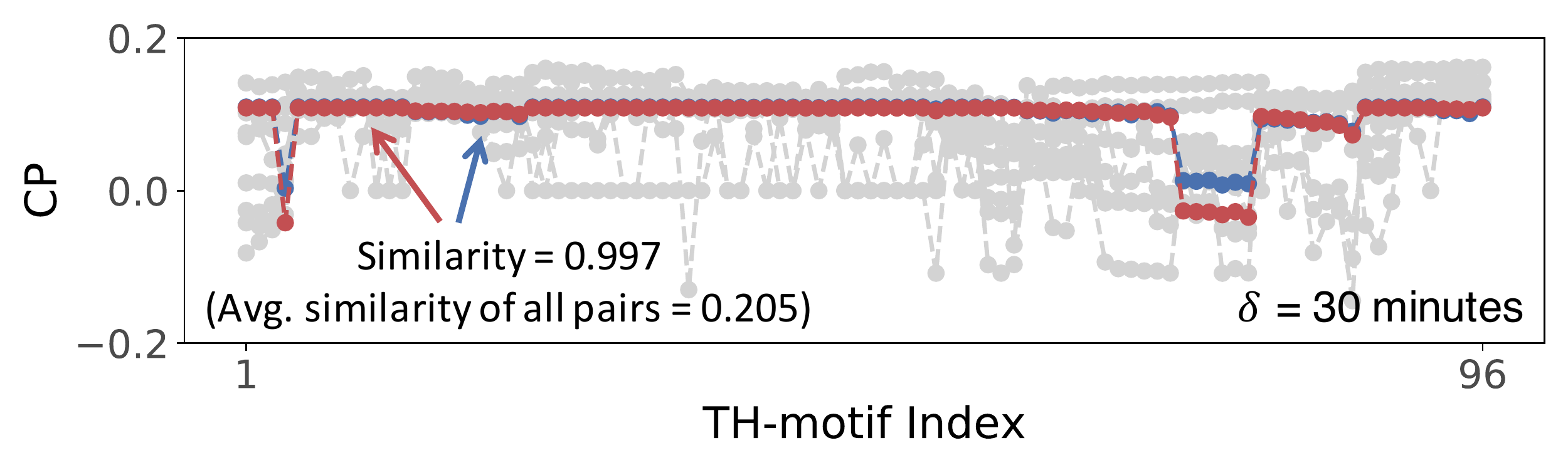}\\
	\vspace{5pt}
	\includegraphics[height=0.55cm]{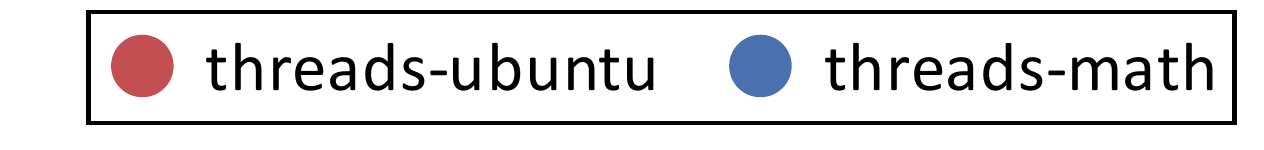}\\
	\vspace{-2.5pt}
	\includegraphics[width=0.95\linewidth]{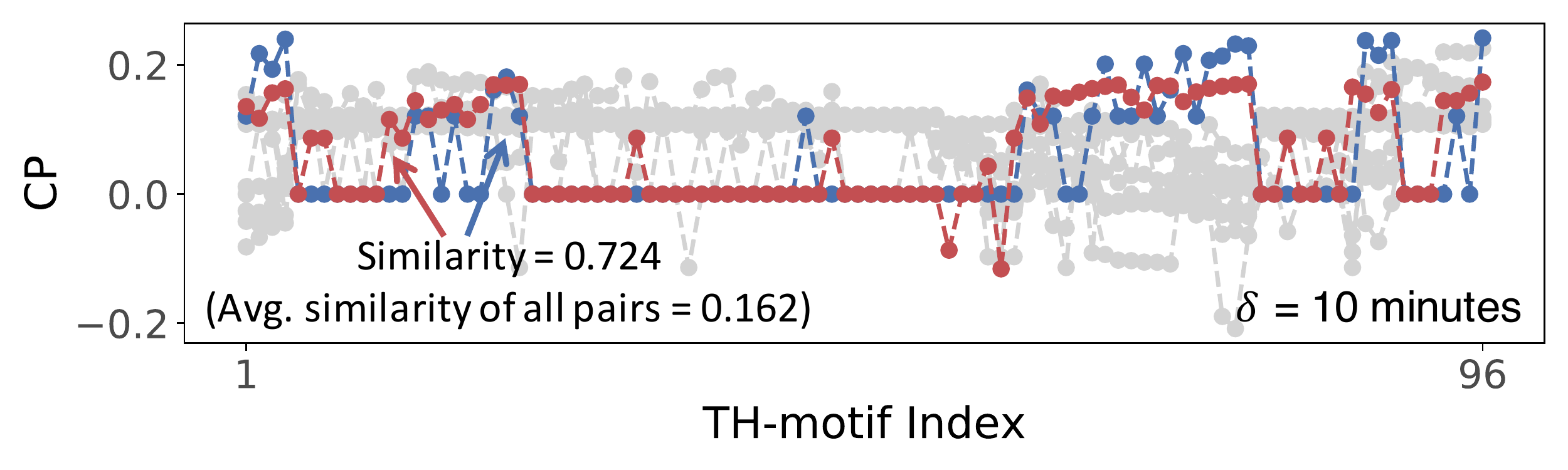}\\
	\vspace{5pt}
	\includegraphics[height=0.55cm]{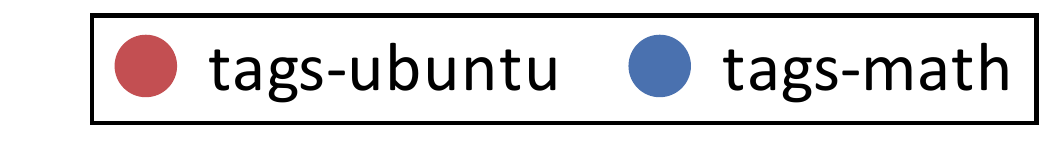}\\
	\vspace{-2.5pt}
	\includegraphics[width=0.95\linewidth]{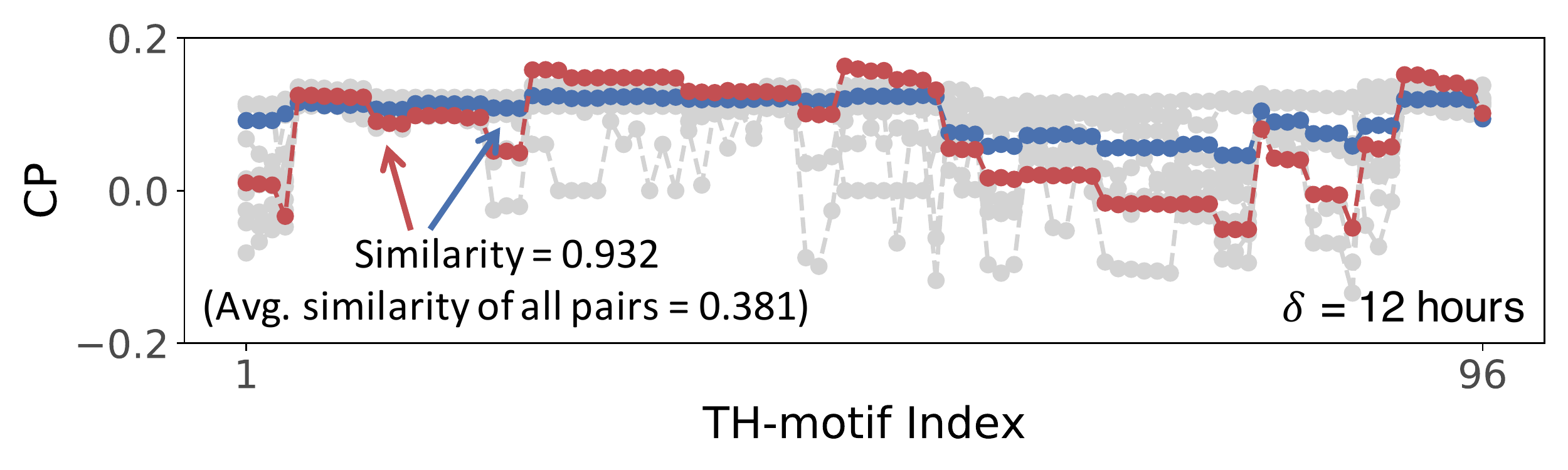}\\
	\vspace{5pt}
	\includegraphics[height=0.55cm]{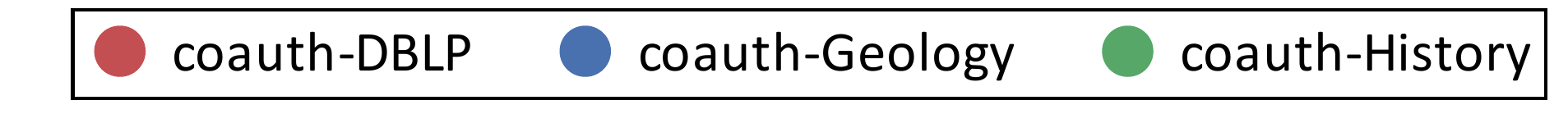}\\
	\vspace{-2.5pt}
	\includegraphics[width=0.95\linewidth]{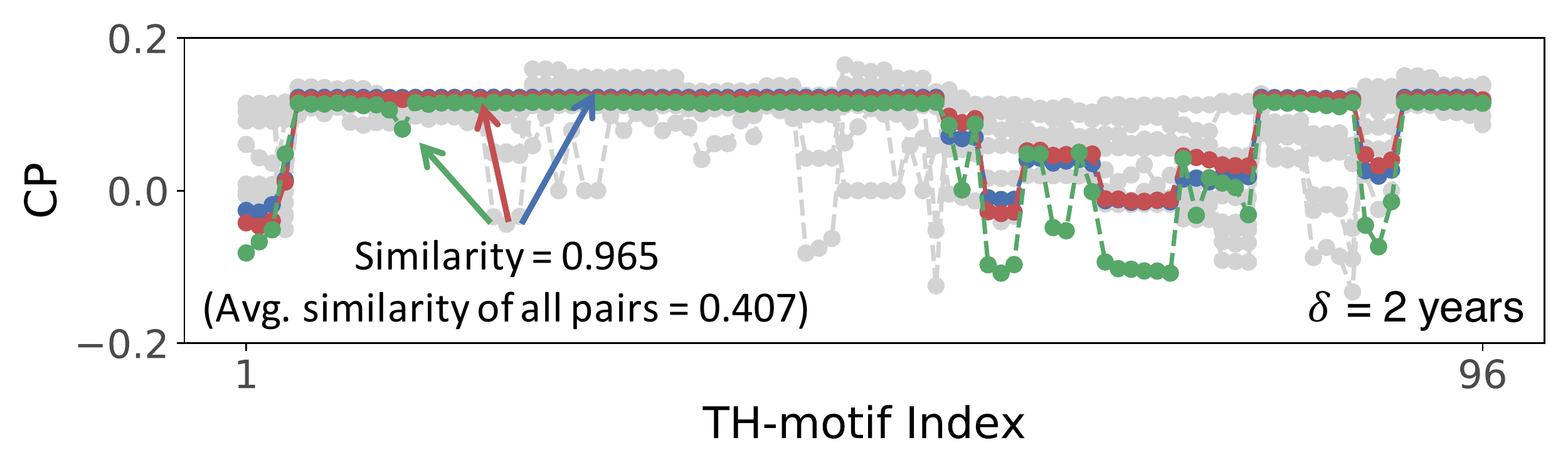}\\
	\vspace{5pt}
	\caption{Characteristic profiles (CPs) (i.e., normalized significance of each \motif) accurately capture the patterns of real-world temporal hypergraphs. 
	The CPs of the temporal hypergraphs from the same domain are similar in terms of the Pearson correlation coefficients, which are the reported numbers, while they are different across domains. Grey lines indicate CPs of the temporal hypergraphs from other domains.\label{fig:cp}}
\end{figure}

In this section, we review experiments to answer Q1-Q4.

\begin{enumerate}[label=Q\arabic*.,leftmargin=*]
    \item \textbf{Discoveries:} Which findings do \motifs bring?
    \item \textbf{Comparison with Static H-motifs:} Are \motifs more informative than static hypergraph motifs \cite{lee2020hypergraph}?
    \item \textbf{Speed \& Efficiency:} How fast and efficient is \adv? 
    \item \textbf{Further Analysis:} Why is \adv fast and efficient? 
\end{enumerate}

We first describe the settings where the experiments are conducted.
Then, we provide some empirical observations using the proposed concepts and algorithms. 
Next, we test the scalability of the methods.
Finally, we provide possible reasons why \adv is efficient based on the observations on real-world temporal hypergraphs.

\begin{table}[t!]
	\begin{center}
		\caption{\label{tab:datasets}Statistics of the 11 real-world hypergraphs from 5 different domains: the number of nodes $|V|$, the number of temporal hyperedges $|\mathcal{E}|$, the number of induced static hyperedges $|E_{\mathcal{E}}|$, and the maximum hyperedge size $\max_{e\in \mathcal{E}}|e|$.}
		\scalebox{0.935}{
			\begin{tabular}{c|c|c|c|c}
				\toprule
				\textbf{Dataset} & $\pmb{|V|}$ & $\pmb{|\mathcal{E}|}$ & $\pmb{|E_{\mathcal{E}}|}$ & $\pmb{\max_{e\in \mathcal{E}}|e|}$\\
				\midrule
				\texttt{email-Enron} & 143 & 10,885 & 1,514 & 37\\
				\texttt{email-Eu} & 986 & 235,263 & 25,148 & 40\\
				\midrule
				\texttt{contact-primary} & 242 & 106,879 & 12,704 & 5\\
				\texttt{contact-high} & 327 & 172,035 & 7,818 & 5\\
				\midrule
				\texttt{threads-ubuntu} & 90,054 & 192,947 & 166,999 & 14\\
				\texttt{threads-math}& 153,806 & 719,792 & 595,749 & 21\\
				\midrule
				\texttt{tags-ubuntu} & 3,021 & 271,233 & 147,222 & 5\\
				\texttt{tags-math} & 1,627 & 822,059 & 170,476 & 5\\
				\midrule
				\texttt{coauth-DBLP} & 1,836,596 & 3,700,681 & 2,467,389 & 280\\
				\texttt{coauth-Geology} & 1,091,979 & 1,591,166 & 1,204,704 & 284\\
				\texttt{coauth-History} & 503,868 & 1,813,147 & 896,062 & 925\\
				\bottomrule %
			\end{tabular}}
	\end{center}
\end{table}

\subsection{Experimental Settings}
\label{sec:experiments:settings}

\smallsection{Machines:}
We conducted all the experiments on a machine with i9-10900K CPU and 64GB RAM.

\smallsection{Implementation:}
We implemented \wsdmshort, \naive, and \adv commonly in C++.

\smallsection{Datasets:}
We use eleven real-world temporal hypergraphs from five different domains. 
Refer to Table~\ref{tab:datasets} for the summarized statistics of the hypergraphs.
We provide the details of each dataset in Appendix B. 
While we assume that timestamps of temporal hyperedges are unique, in some dataset, this may not hold.
In such cases, we randomly order the temporal hyperedges whose timestamps are identical.

\subsection{Q1. Discoveries}

In this subsection, we present several observations that \shorts reveal in the $11$ real-world hypergraphs. 
\shorts provide a new perspective in analyzing temporal hypergraphs.

\smallsection{Obs~1. Real hypergraphs are not `random':}
For an accurate characterization, we compare the number of instances of \motifs in real-world temporal hypergraphs against that in randomized ones. 
To this end, we randomize the real-world temporal hypergraph using HyperCL~\cite{lee2021hyperedges}, a random hypergraph generator which preserves node degrees and hyperedge sizes.
Once the randomized hypergraph is generated, we randomly assign the timestamps of its temporal hyperedges.
In Fig.~\ref{fig:counts}, we compare the distribution of the number of instances of each \motif in real-world temporal hypergraphs and those in randomized ones.
The distributions are clearly different, and the total number of instances is greater in real-world hypergraphs than in random hypergraphs.
Specifically, the total number of \motifs' instances in real-world hypergraphs are $6.42\times$, $1.44\times$, $46.69\times$, $4.30\times$ of that in randomized hypergraphs in \texttt{email-Eu}, \texttt{contact-primary}, \texttt{threads-math}, and \texttt{tags-ubuntu}, respectively.

\smallsection{Obs~2. \motifs distinguish domains:}
Network motifs have demonstrated their power to distinguish graphs based on their domains. 
In addition, the count distributions of h-motifs in static hypergraphs are particularly similar between domains but different across domains. 
To confirm that temporal h-motifs also possess such distinguishing power, we obtain the characteristic profile (CP) of each hypergraph, a normalized 96 dimensional vector of concatenation of relative significance of each temporal h-motif, as suggested in \cite{lee2020hypergraph}.
As seen in Fig.~\ref{fig:cp}, CPs accurately capture patterns of real-world temporal hypergraphs. 
That is, while CPs of the temporal hypergraphs from the same domain are similar, they are different across domains. 
These results support that \motifs play a key role in capturing structural and temporal patterns of real-world temporal hypergraphs. 

\begin{figure}[t]
	\vspace{-6pt}
	\centering
	\includegraphics[width=0.45\textwidth]{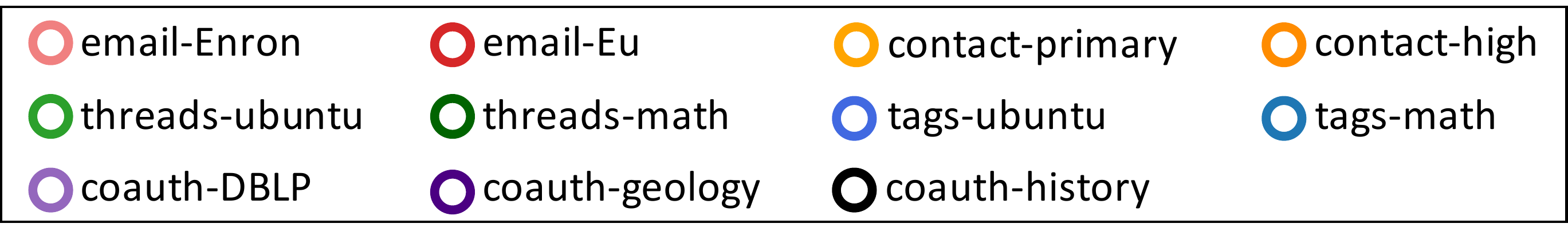}
	\begin{subfigure}[b]{.16\textwidth}
    	\includegraphics[width=0.97\linewidth]{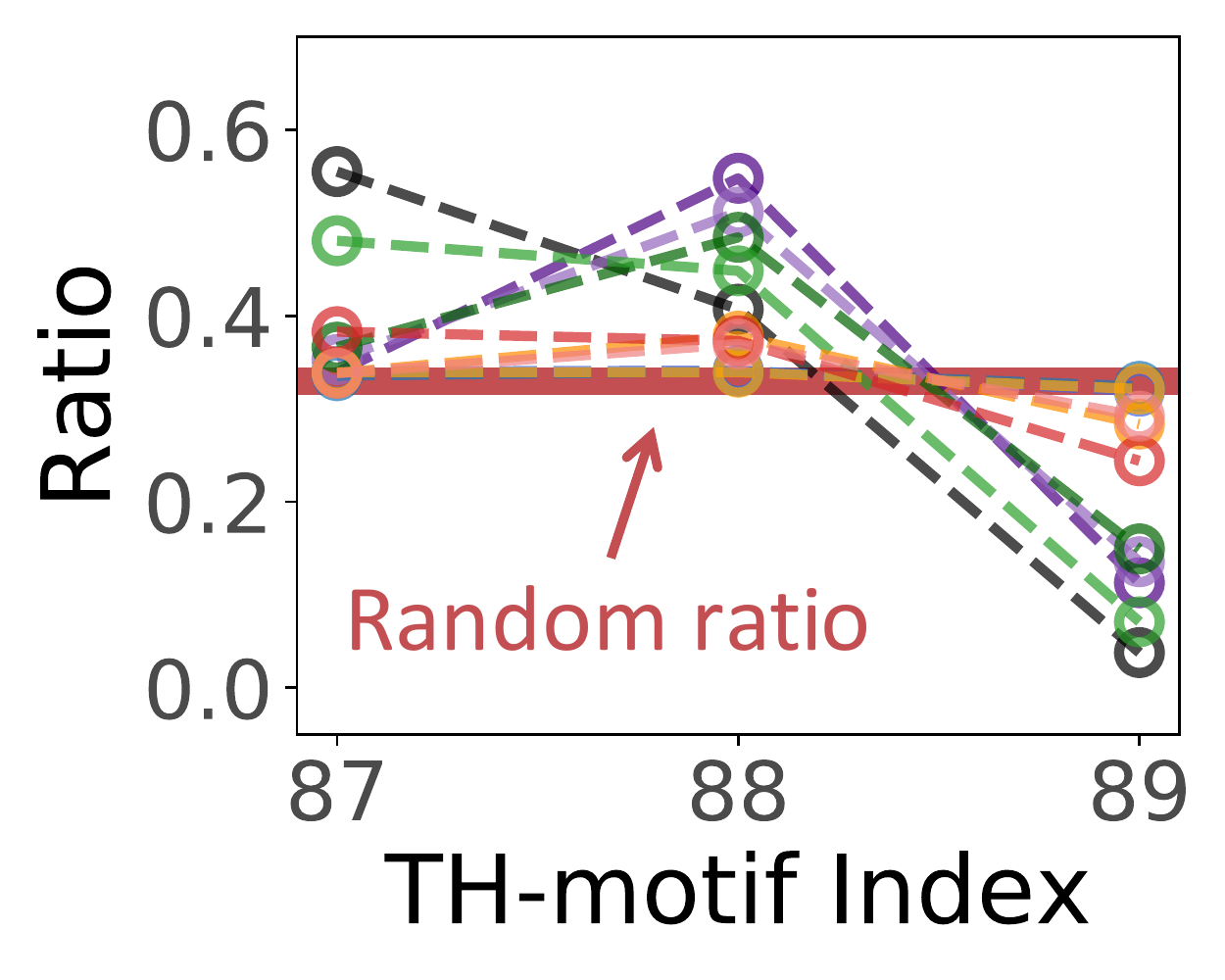}
    	\vspace{-1pt}
    	\caption{\motifs 87-89}
	\end{subfigure}
	\hspace{-6pt}
	\begin{subfigure}[b]{.16\textwidth}
    	\includegraphics[width=0.97\linewidth]{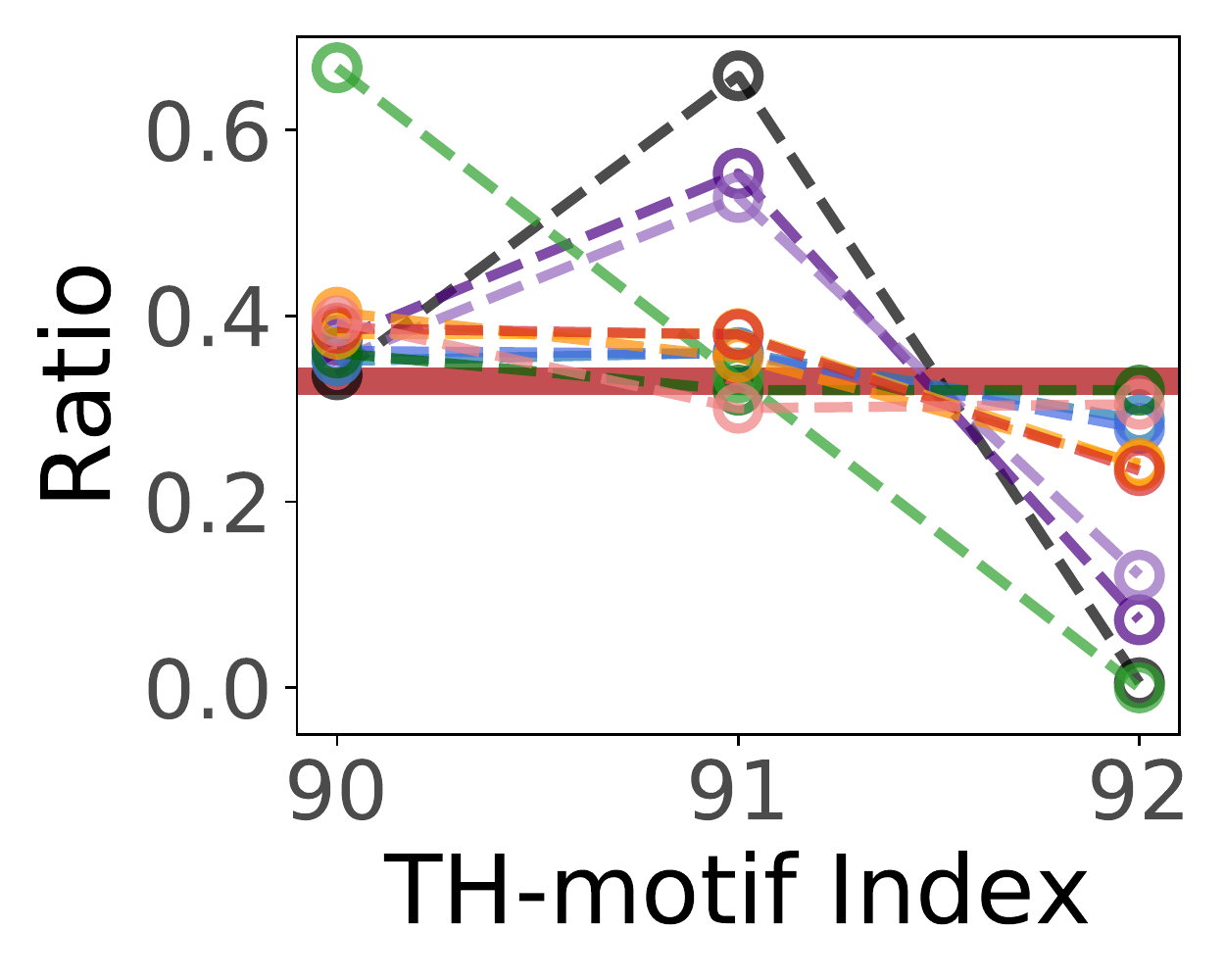}
    	\vspace{-1pt}
    	\caption{\motifs 90-92}
	\end{subfigure}
	\hspace{-6pt}
	\begin{subfigure}[b]{.16\textwidth}
    	\includegraphics[width=0.97\linewidth]{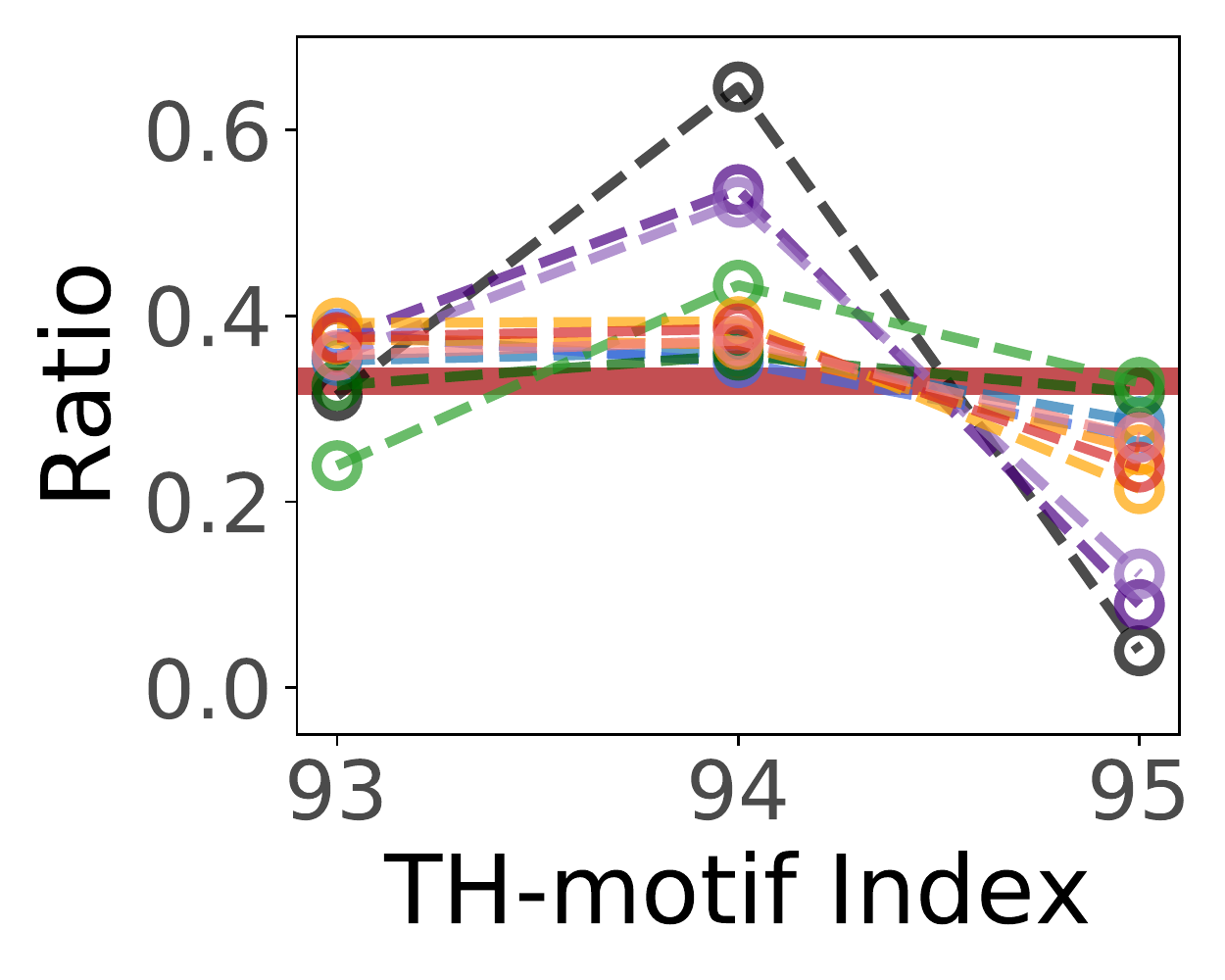}
    	\vspace{-1pt}
    	\caption{\motifs 93-95}
	\end{subfigure}
	\caption{The number of instances of nine \pair \motifs depend on the ordering of the hyperedges. The ratio of the occurrences of \motifs 89, 92, and 95 are significantly low compared to the other \motifs with same structures. \label{fig:pair}}
\end{figure}

\smallsection{Obs~3. Orders of hyperedges matter:}
\motifs are asymmetric with respect to the arrival order of the temporal hyperedges, and thus instances that are indistinguishable with static h-motifs can be categorized as different \motifs. 
We are interested in how the orders of the hyperedges affect the occurrences of \motifs, and to this end, we statistically investigate nine \pair ones, ranging from \motif 87 to 95. 
\motifs in each triple, \motifs $87-89$, \motifs $90-92$, and \motifs $93-95$ share the same structural pattern and are distinguished by the orders of the hyperedges.
Consider an instance $\langle e_i,e_j,e_k \rangle$ of the \pair \motif. 
The \pair \motifs, by definition, consist of a pair of duplicated hyperedges and thus enables three different orderings \textbf{O1}: $\tilde{e}_i = \tilde{e}_j \neq \tilde{e}_k$, \textbf{O2}: $\tilde{e}_i \neq \tilde{e}_j = \tilde{e}_k$, and \textbf{O3}: $\tilde{e}_i \neq \tilde{e}_j \neq \tilde{e}_k$, $\tilde{e}_i = \tilde{e}_k$,.
In \textbf{O1} and \textbf{O2}, duplicated temporal hyperedges occur consecutively, whereas in \textbf{O3}, the first and last hyperedges are duplicated. 
\motifs 87, 90, and 93 are \textbf{O1}, \motifs 88, 91, and 94 are \textbf{O2}, and \motifs 89, 92, and 95 are \textbf{O3}.
As seen in Fig.~\ref{fig:pair}, this difference indeed affect the occurrences of the \motifs in real-world temporal hypergraphs. The ratio of the \motifs whose ordering is \textbf{O3} are significantly small, compared to that of \textbf{O1} and \textbf{O2}.
That is, duplicated temporal hyperedges tend to occur in a short time and thus affect the count distributions of \motifs.


\begin{figure*}[t]
	\vspace{-4mm}
	\centering
	\includegraphics[width=0.09\textwidth]{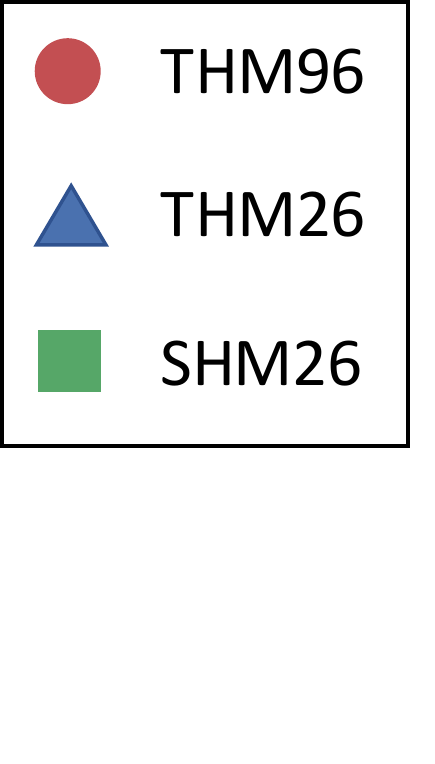}
	\hspace{3pt}
	\begin{subfigure}[b]{.205\textwidth}
    	\includegraphics[width=0.95\linewidth]{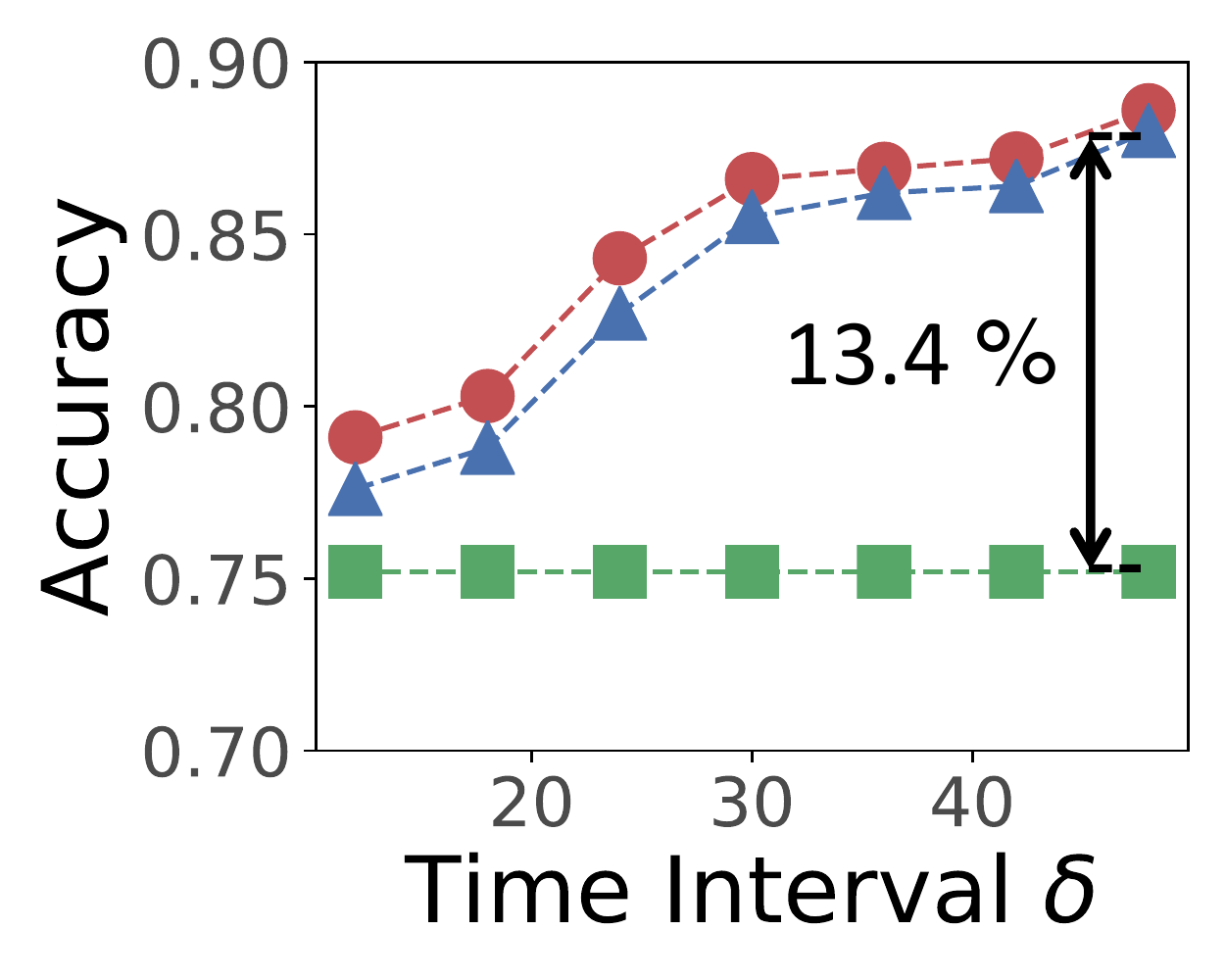}
    	\vspace{-2pt}
    	\caption{\texttt{email-Enron}}
	\end{subfigure}
	\hspace{3pt}
	\begin{subfigure}[b]{.205\textwidth}
    	\includegraphics[width=0.95\linewidth]{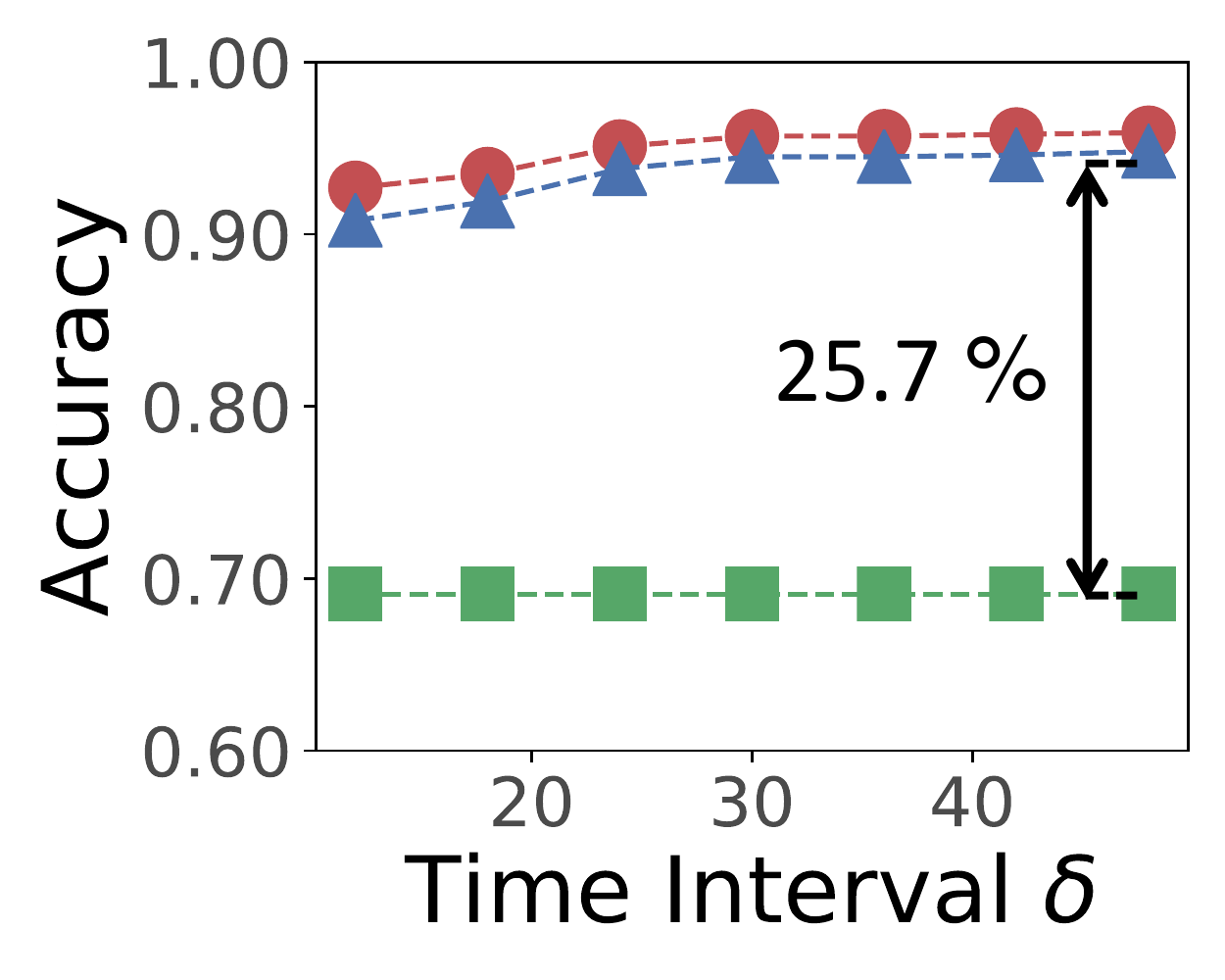}
    	\vspace{-2pt}
    	\caption{\texttt{email-Eu}}
	\end{subfigure}
	\hspace{3pt}
	\begin{subfigure}[b]{.205\textwidth}
    	\includegraphics[width=0.95\linewidth]{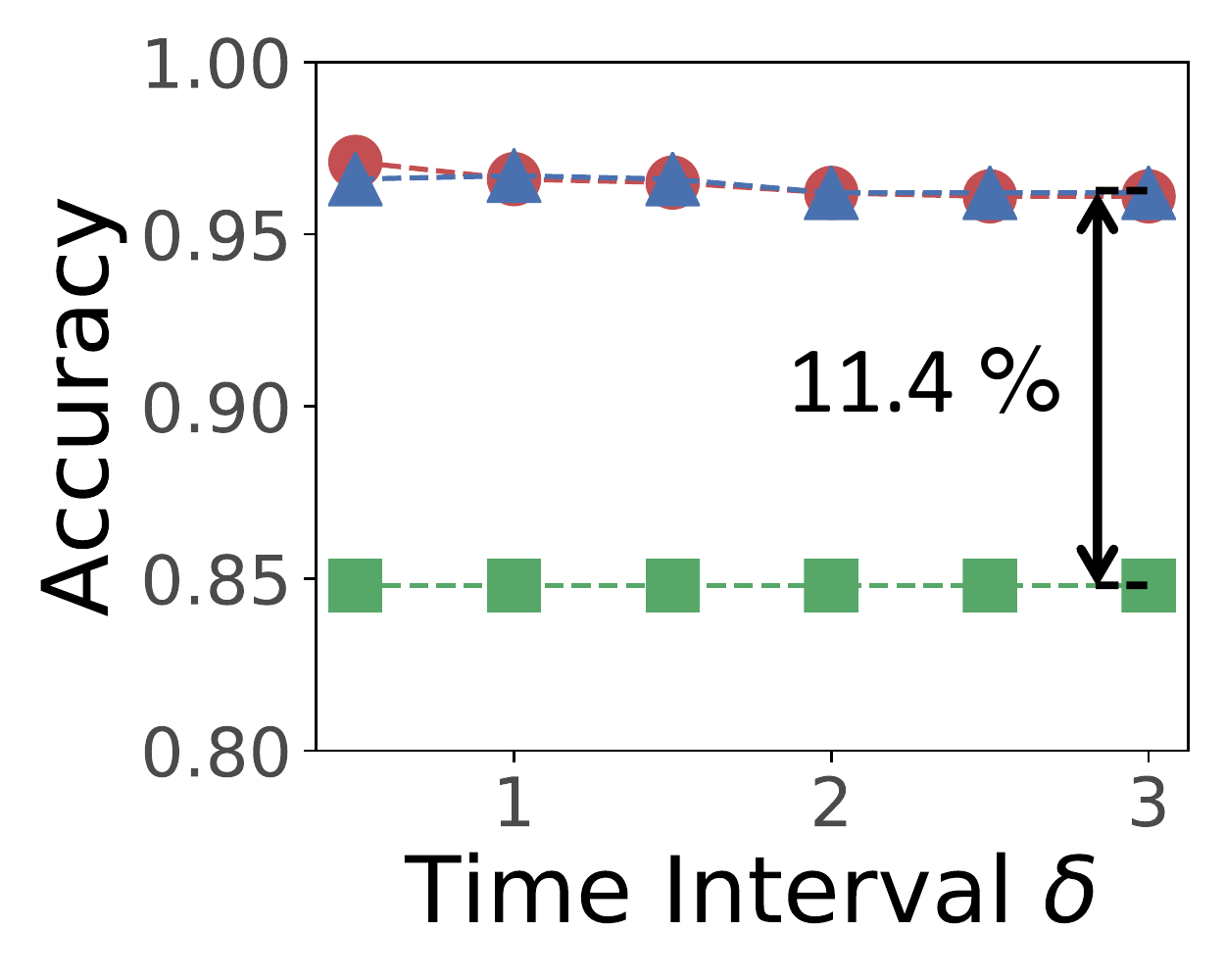}
    	\vspace{-2pt}
    	\caption{\texttt{contact-primary}}
	\end{subfigure}
	\hspace{3pt}
	\begin{subfigure}[b]{.205\textwidth}
    	\includegraphics[width=0.95\linewidth]{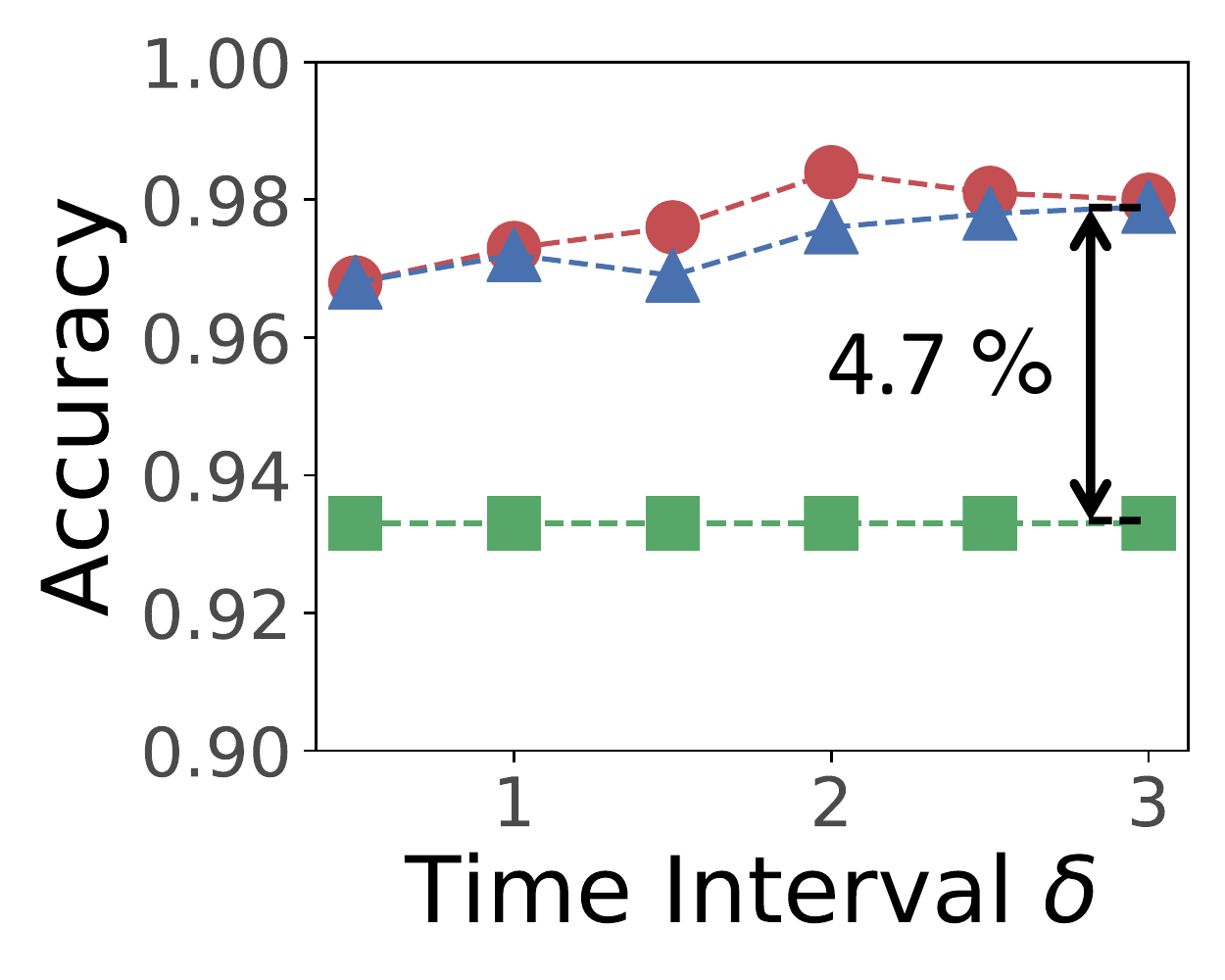}
    	\vspace{-2pt}
    	\caption{\texttt{contact-high}}
	\end{subfigure}
	\caption{\motifs provide informative features of temporal hyperedges. THM96 and THM26, which use the counts of \motifs' instance as features, are more accurate than SHM26, which uses the counts of static h-motifs' instances, in predicting future temporal hyperedges. Results in small datasets where the instances of static h-motifs can be exactly counted are reported. \label{fig:prediction}}
\end{figure*}

\subsection{Q2. Comparison with Static H-motifs}

In this subsection, we demonstrate the usefulness of \shorts.  We compare \shorts and \statics as inputs features for a hyperedge prediction task.

\smallsection{Obs~4. \motifs help predict future hyperedges:}
To verify the usefulness of temporal h-motifs, we consider the problem of hyperedge prediction, a binary classification problem of predicting whether the given hyperedge is true or not. 
Given a temporal hypergraph $T=(V,\mathcal{E})$, we generate a set $\mathcal{E}'$ of fake hyperedges, whose size is equal to the true one (i.e., $|\mathcal{E}|=|\mathcal{E}'|$), using HyperCL~\cite{lee2021hyperedges}, which preserves the degrees of the nodes and the sizes are equal to the true ones.
The timestamps of the fake hyperedges are randomly assigned.
We sort the entire temporal hyperedges $\mathcal{E}\cup \mathcal{E}'$ based on their timestamps and split into train and test sets in a ratio 8:2.
Then we train a logistic regression classifier using the train set with following three different features of each temporal hyperedge:

\begin{itemize}[leftmargin=*]
    \item \textbf{THM96 ($\in \mathbb{R}^{96}$):} Each dimension represents the number of instances of \motifs that contain the hyperedge.
    \item \textbf{THM26 ($\in \mathbb{R}^{26}$):} The 26 \motifs whose occurrences have the highest variance are selected.
    \item \textbf{SHM26 ($\in \mathbb{R}^{26}$):} Each dimension represent the number of instances of static h-motifs that contain the hyperedge. Temporal information is ignored.
\end{itemize}

\noindent As seen in Fig.~\ref{fig:prediction}, THM96 and THM26, which are based on the \motifs counts, are more accurate than STM26.
While h-motifs only represent structural patterns, \motifs incorporate temporal information in addition to them, and thus they are more informative.

\begin{figure*}[t]
	\vspace{-7mm}
	\centering
	\includegraphics[width=0.30\textwidth]{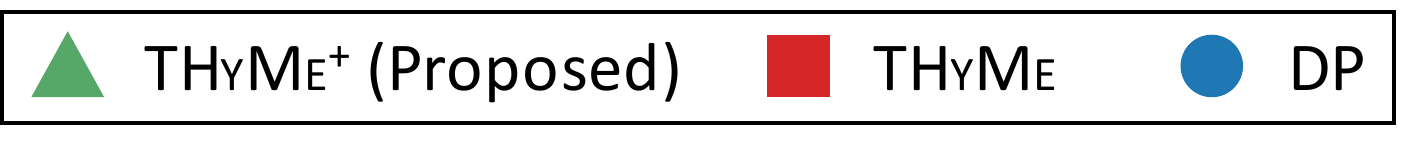}\\
	\begin{subfigure}[b]{.19\textwidth}
    	\includegraphics[width=0.95\linewidth]{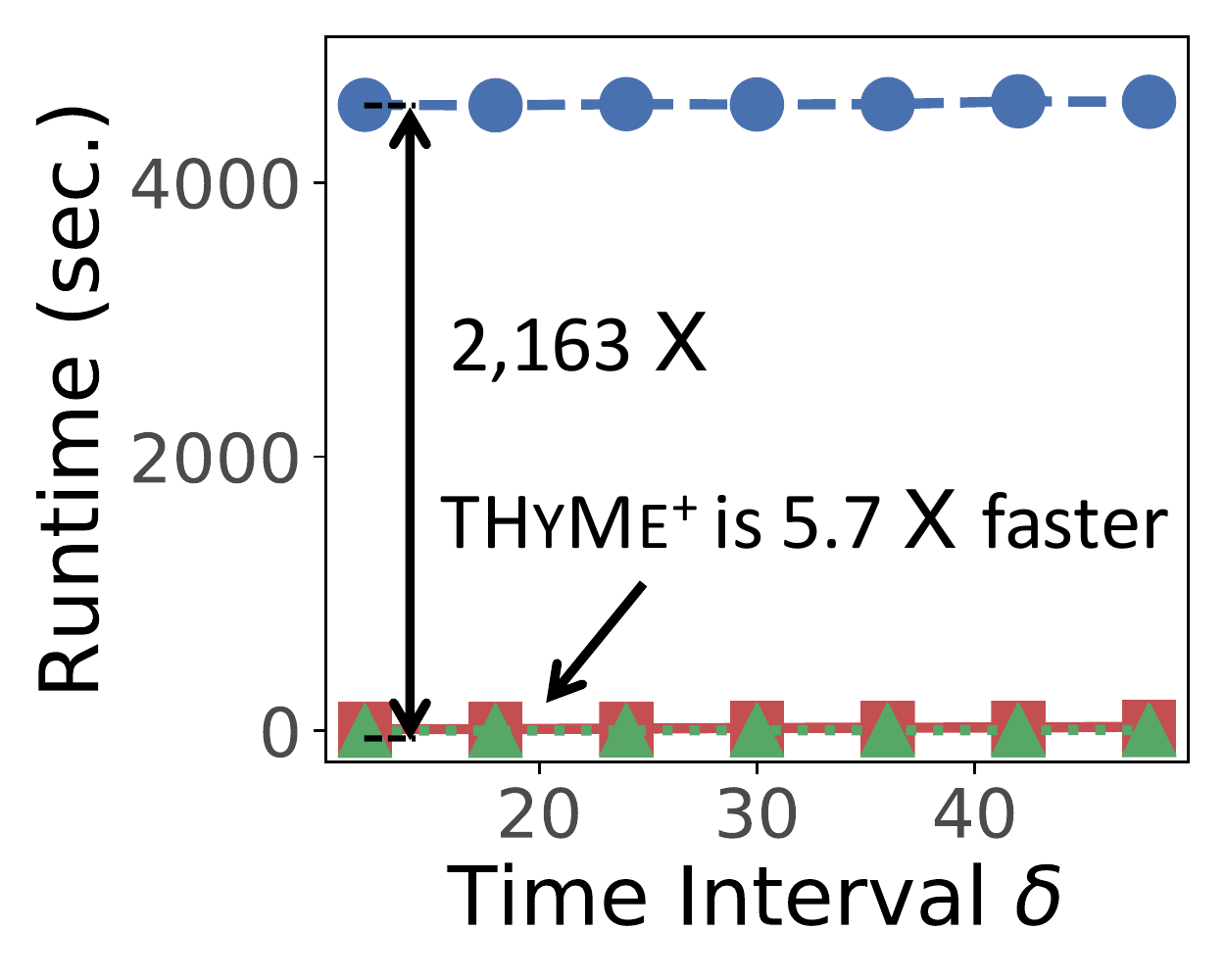}
    	\vspace{-2pt}
    	\caption{\texttt{email-Eu}}
	\end{subfigure}
	\hfill
	\begin{subfigure}[b]{.19\textwidth}
    	\includegraphics[width=0.95\linewidth]{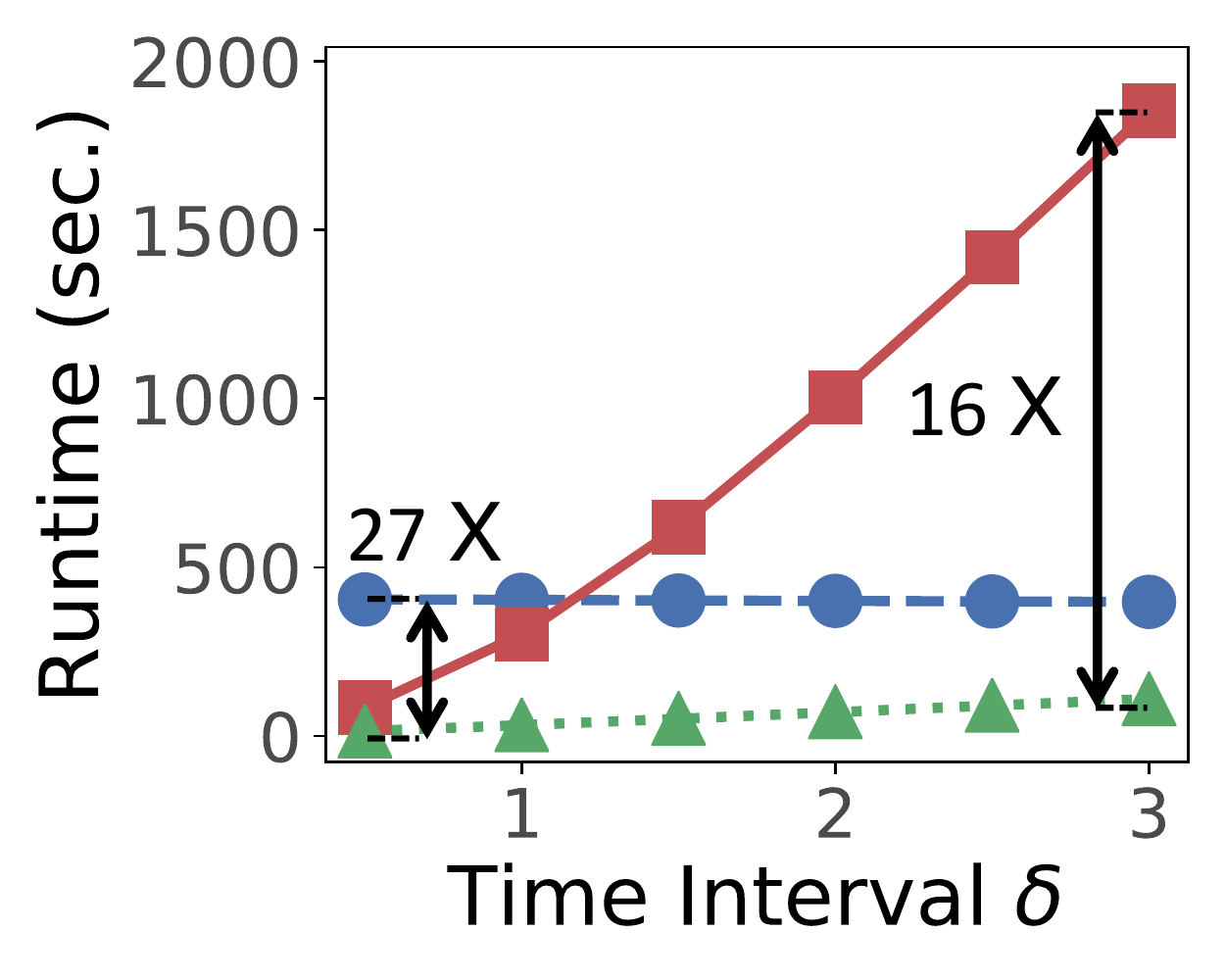}
    	\vspace{-2pt}
    	\caption{\texttt{contact-primary}}
	\end{subfigure}
	\hfill
	\begin{subfigure}[b]{.19\textwidth}
    	\includegraphics[width=0.95\linewidth]{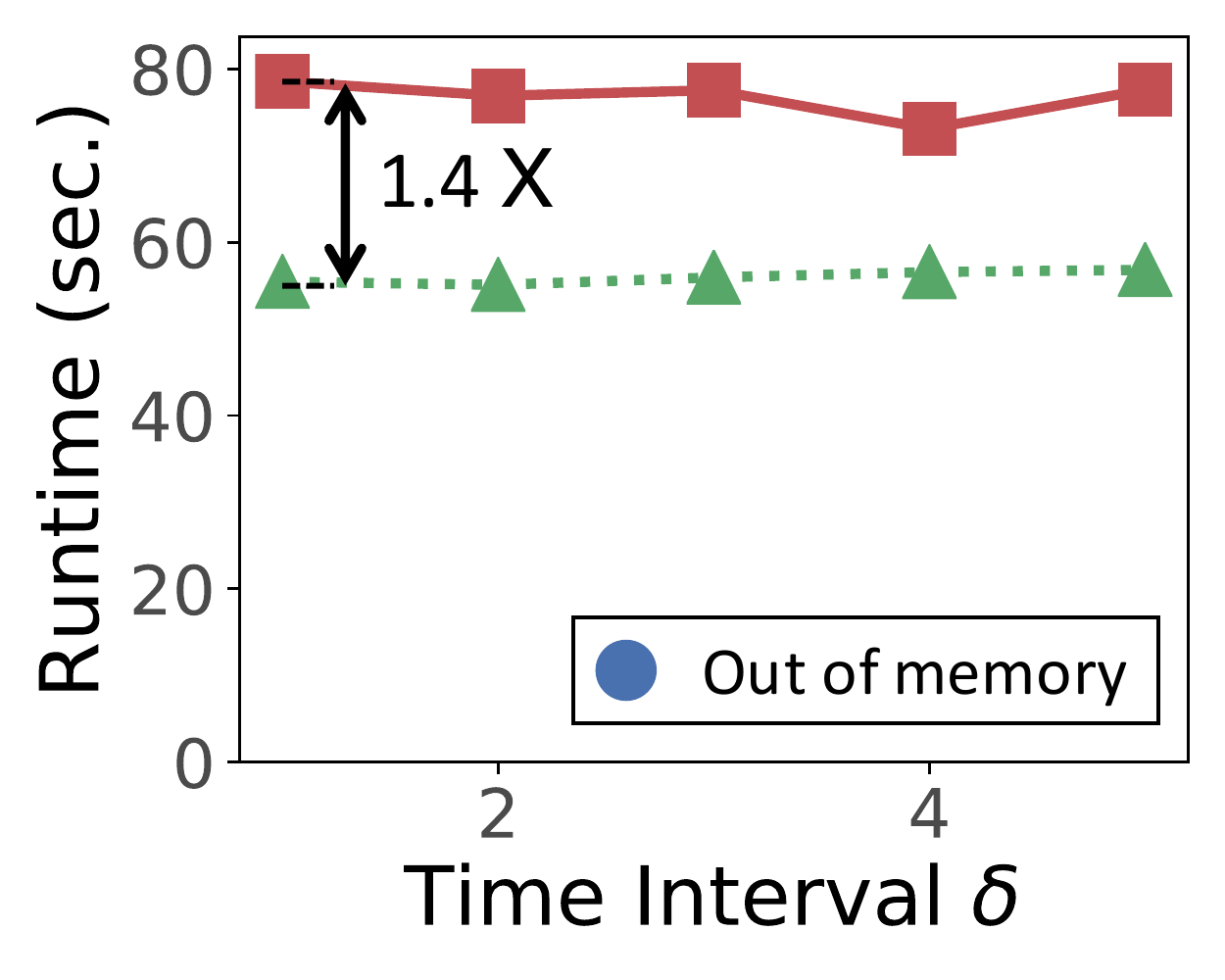}
    	\vspace{-2pt}
    	\caption{\texttt{threads-math}}
	\end{subfigure}
	\hfill
	\begin{subfigure}[b]{.19\textwidth}
    	\includegraphics[width=0.95\linewidth]{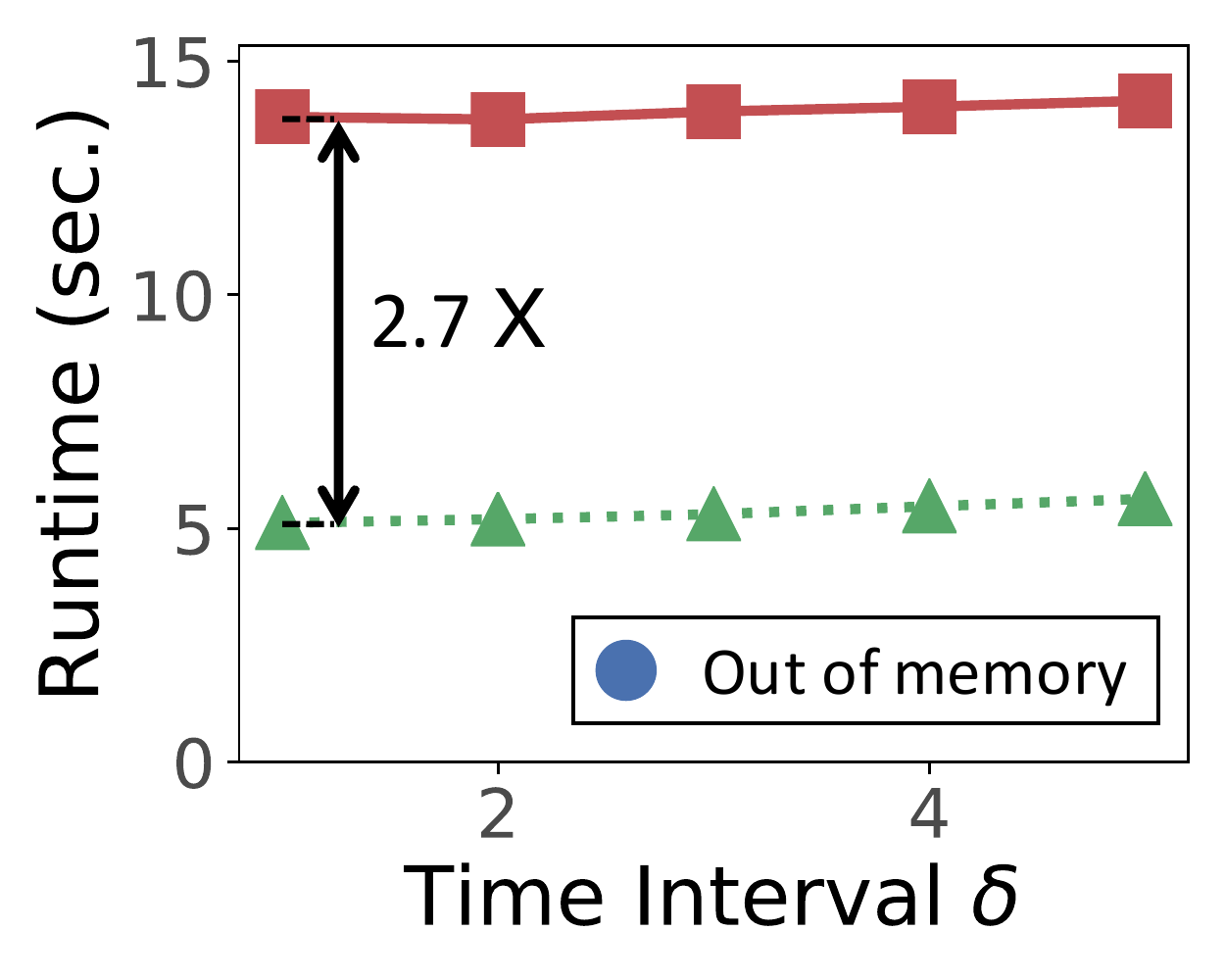}
    	\vspace{-2pt}
    	\caption{\texttt{tags-ubuntu}}
	\end{subfigure}
	\hfill
	\begin{subfigure}[b]{.19\textwidth}
    	\includegraphics[width=0.95\linewidth]{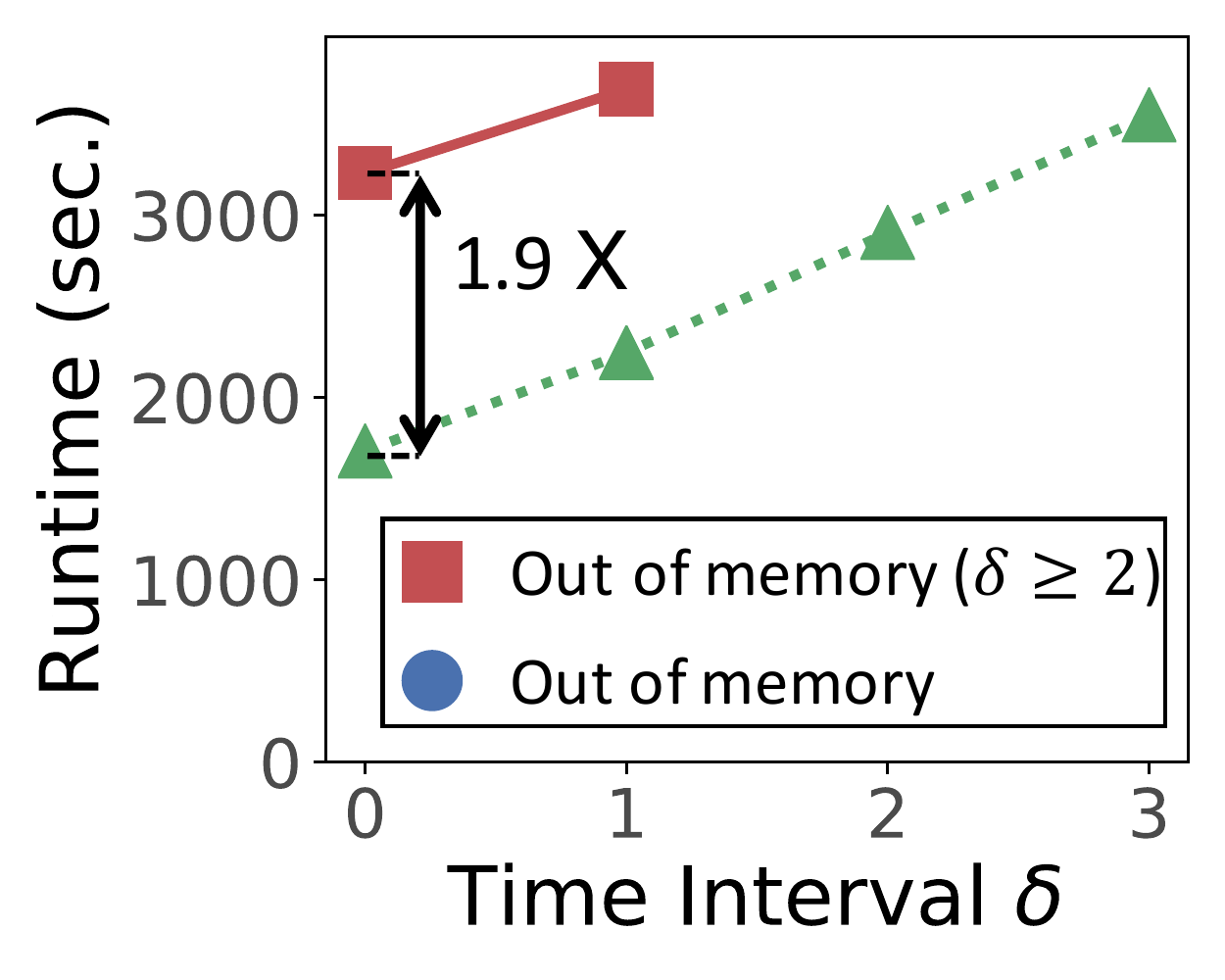}
    	\vspace{-2pt}
    	\caption{\texttt{coauth-DBLP}}
	\end{subfigure}
	\caption{\adv is faster and more space efficient than \wsdmshort and \naive. We provide the full results in \cite{online2021appendix}.\label{fig:runtime}}
\end{figure*}

\subsection{Q3. Speed and Efficiency}
We evaluate the speed and efficiency of the proposed algorithms \wsdmshort, \naive, and \adv.
As seen in Fig.~\ref{fig:runtime}, while \wsdmshort and \naive run out of memory in some datasets or with particular $\delta$ values, \adv is fast and space efficient enough in all considered settings.
Specifically, \adv is up to $2,163 \times$ faster than \wsdmshort and $16 \times$ faster than \naive.
As described in Section~\ref{sec:method}, \adv maintains a small projected graph $Q$ and thus reduces enumeration over the instances in $Q$. In the next subsection, we provide empirical findings that support the effectiveness of \adv.

\subsection{Q4. Further Analysis\label{sec:experiments:analysis}}
Why is \adv faster and more space efficient compared to \wsdmshort and \naive?
What properties of real-world temporal hypergraphs make \adv efficient?
To answer these questions, we examine structural and temporal patterns of temporal hyperedges in real-world temporal hypergraphs and summarize common properties observed as follows.

\begin{itemize}[leftmargin=*]
    \item \textbf{(Obs. 5) Repetitive behavior:} Duplicated temporal hyperedges tend to appear repeatedly, and the distribution of the numbers of repetitions is heavy-tailed. 
    \item \textbf{(Obs. 6) Temporally locality:} Future temporal hyperedges are more likely to repeat recent hyperedges than older ones.
\end{itemize}

\begin{figure}[t]
	\vspace{-2mm}
	\centering
	\begin{subfigure}[b]{.155\textwidth}
    	\includegraphics[width=0.99\linewidth]{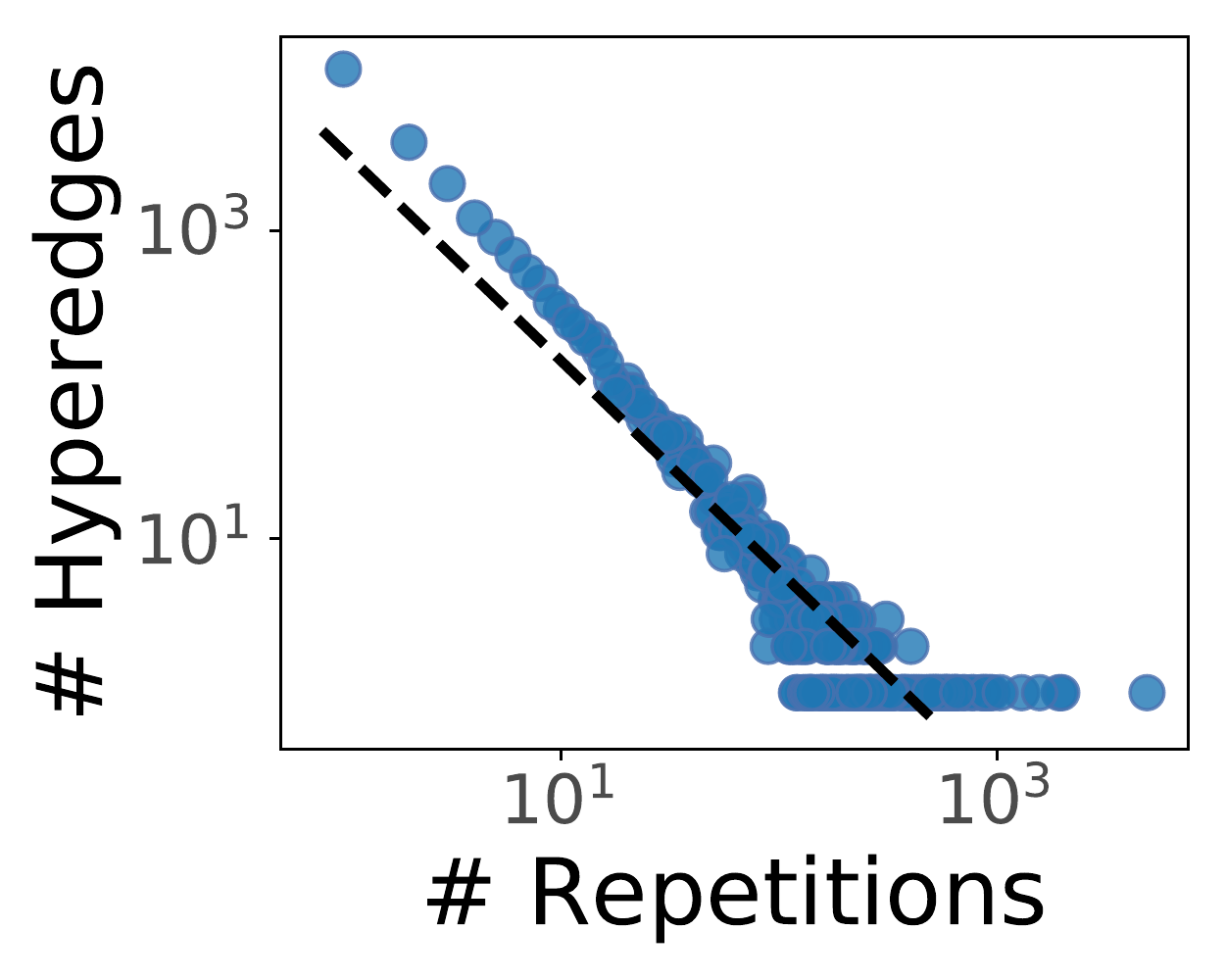}
    	\vspace{-15pt}
    	\caption{\small{\texttt{email-Eu}}}
	\end{subfigure}
	\begin{subfigure}[b]{.155\textwidth}
    	\includegraphics[width=0.99\linewidth]{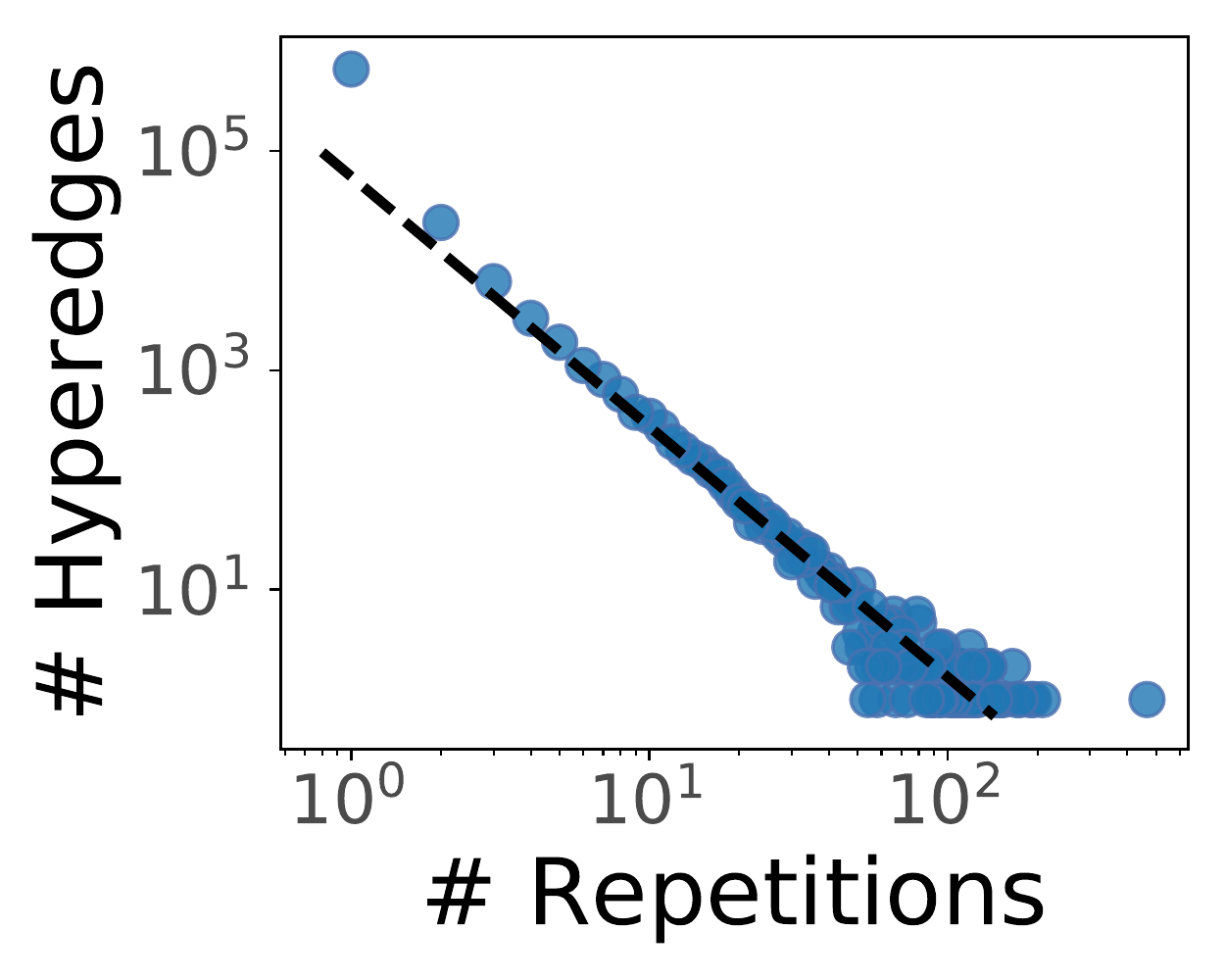}
    	\vspace{-15pt}
    	\caption{\small{\texttt{threads-math}}}
	\end{subfigure}
	\begin{subfigure}[b]{.155\textwidth}
    	\includegraphics[width=0.99\linewidth]{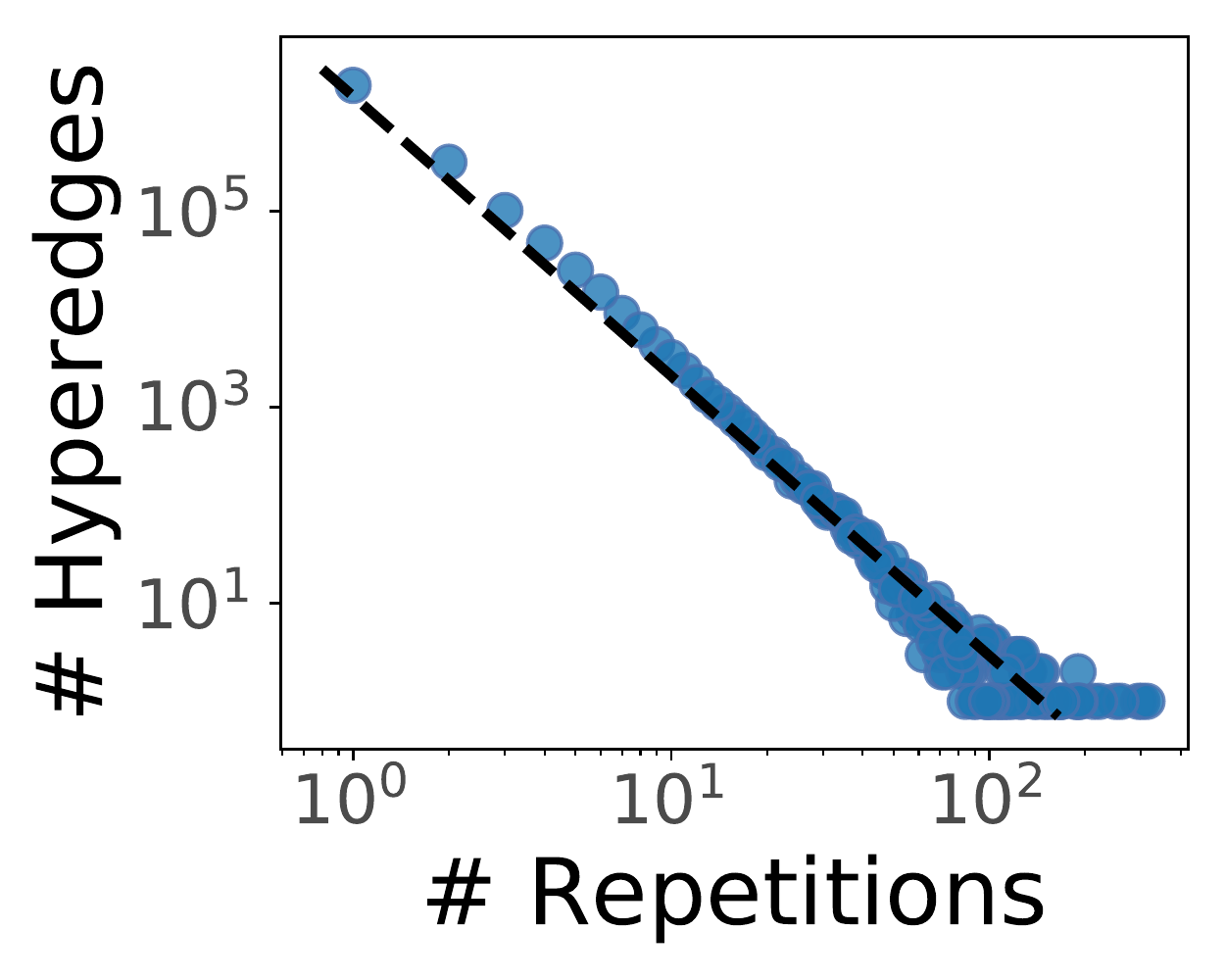}
    	\vspace{-15pt}
    	\caption{\small{\texttt{coauth-DBLP}}}
	\end{subfigure}
	\caption{Temporal hyperedges in real-world hypergraphs are repetitive. Temporal hyperedges appear repetitively and the number of repetitions follow a near power-law distribution. This tendency is found consistently across all datasets \cite{online2021appendix}. 
	\label{fig:powerlaw}}
\end{figure}

\smallsection{Obs. 5. Repetitive behavior:}
We first investigate the repeating patterns (i.e., duplication) of temporal hyperedges in real-world temporal hypergraphs. 
As seen in Table~\ref{tab:datasets}, the number of induced hyperedges ($|E_{\mathcal{E}}|$) is significantly smaller than that of temporal hyperedges ($|\mathcal{E}|$), implying that temporal hyperedges are frequently repeated. 
Surprisingly, in \texttt{contact-high} dataset, the number of induced hyperedges is only $4.5\%$ of that of temporal hyperedges, implying that most temporal hyperedges consist of predefined set of nodes.
Note that due to the flexibility of hyperedge sizes, a hyperedge can be generated from $O(2^{|V|})$, and thus is extremely unlikely to repeat the exact set of nodes.
In addition, we discover that the distributions of hyperedge repetitions in real-world temporal hypergraphs are generally heavy-tailed and close to power-law distributions, as seen in Fig.~\ref{fig:powerlaw}.
We support this claim by fitting the distributions to representative heavy-tailed distributions in \cite{online2021appendix}.

\begin{figure}[t]
	\vspace{-2mm}
	\centering
	\includegraphics[width=0.36\textwidth]{FIG/locality_legend.pdf}
	\begin{subfigure}[b]{.155\textwidth}
    	\includegraphics[width=0.99\linewidth]{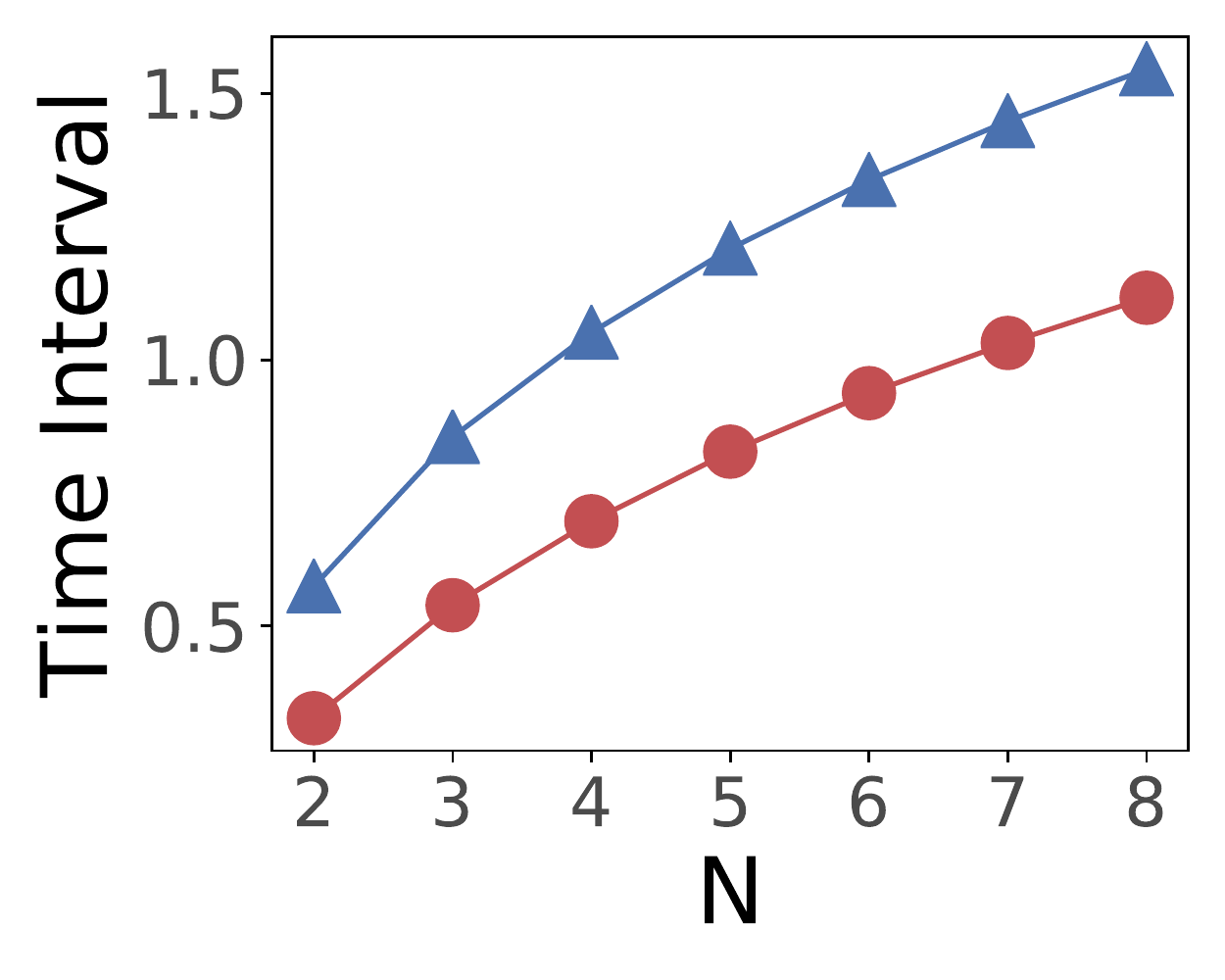}
    	\vspace{-15pt}
    	\caption{\small{\texttt{email-Eu}}}
	\end{subfigure}
	\begin{subfigure}[b]{.155\textwidth}
    	\includegraphics[width=0.99\linewidth]{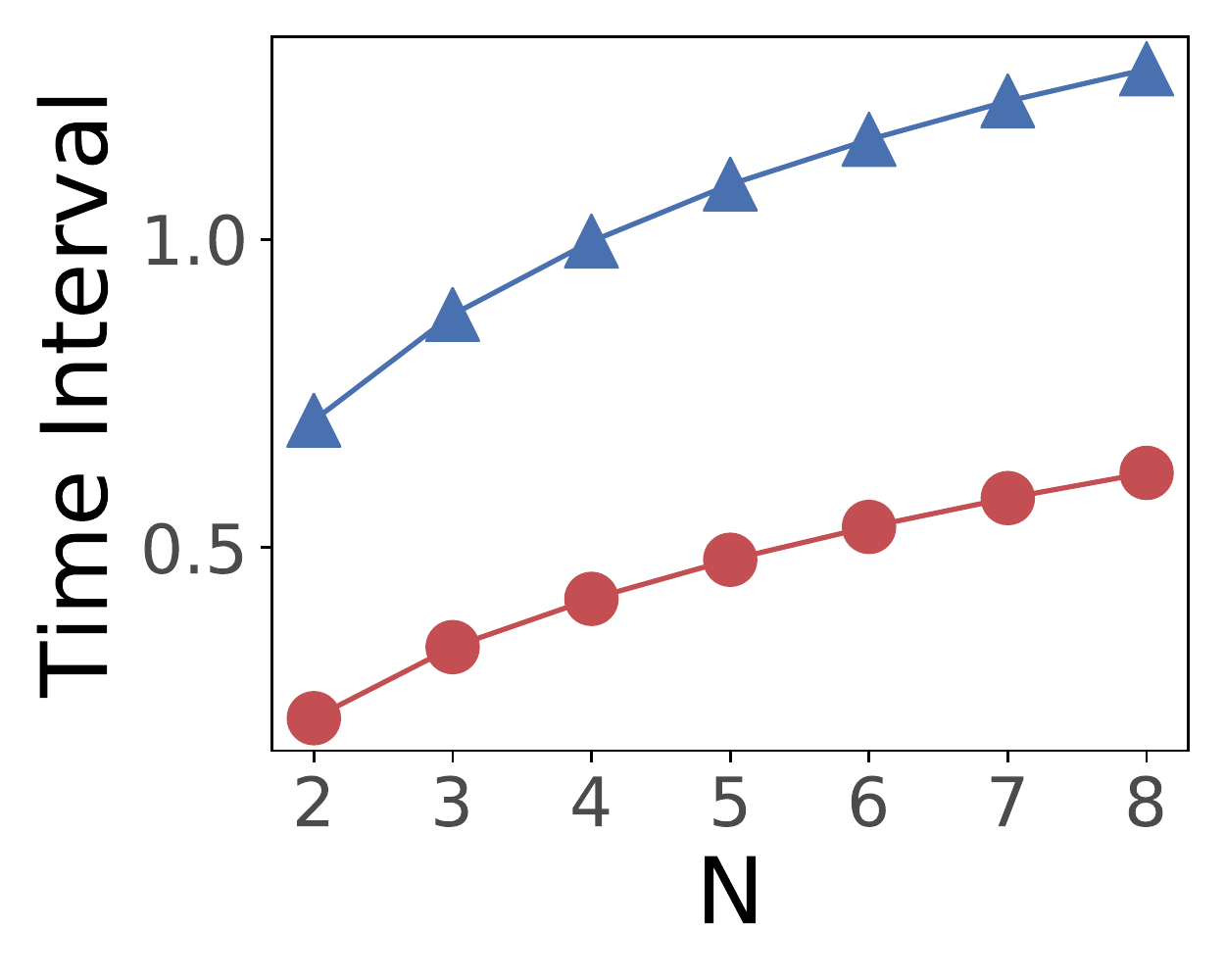}
    	\vspace{-15pt}
    	\caption{\small{\texttt{threads-math}}}
	\end{subfigure}
	\begin{subfigure}[b]{.155\textwidth}
    	\includegraphics[width=0.99\linewidth]{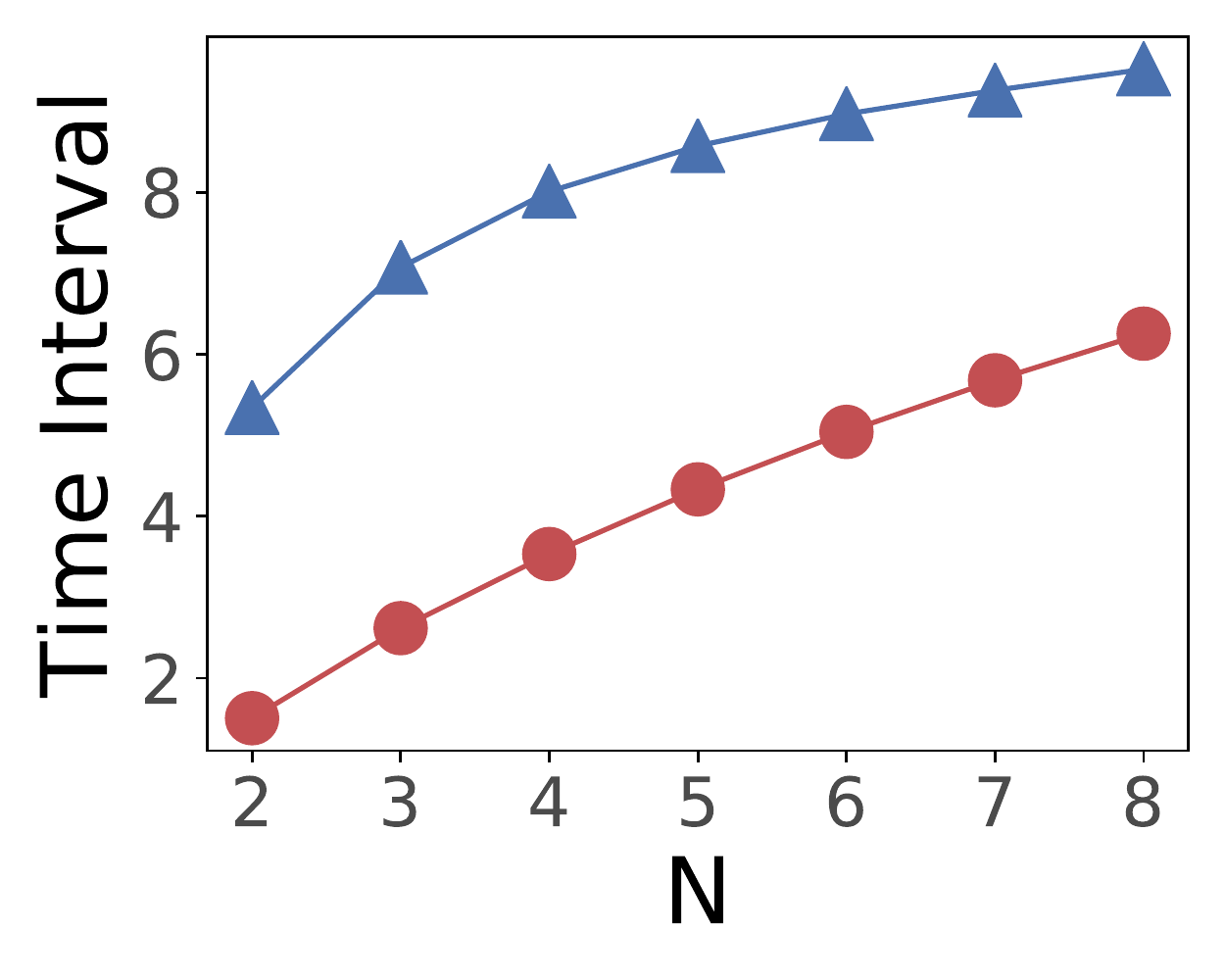}
    	\vspace{-15pt}
    	\caption{\small{\texttt{coauth-DBLP}}}
	\end{subfigure}
	\caption{Temporal hyperedges in real-world hypergraphs are temporally local. The time intervals of $N$ consecutive duplicated temporal hyperedges is shorter in real-world temporal hypergraphs than in randomized hypergraphs. The units of time intervals in \texttt{coauth-DBLP} is years, and the others are hours. We provide the full results in \cite{online2021appendix}.\label{fig:locality}}
\end{figure}

\smallsection{Obs. 6. Temporal locality:}
Now that we have observed the structural behaviors of the temporal hyperedges, we turn our attention to the temporal aspect. The temporal locality of temporal hyperedges is the tendency that recent hyperedges are more likely to be repeated in the near future than the older ones.
To show the temporal locality, we investigate the time intervals of the $N$ consecutive identical temporal hyperedges, i.e., the time it takes for a hyperedge to be repeated $N$ times.
Fig.~\ref{fig:locality} shows the average time intervals of all the hyperedges in the real-world hypergraphs and randomly shuffled hypergraphs, where timestamps of the hyperedges are randomly shuffled while preserving the underlying structure.
In every dataset, the time intervals within $N$ consecutive hyperedges are shorter in real-world hypergraphs than in randomized ones. That is, future hyperedges are more likely to repeat the recent hyperedges than older ones.


\smallsection{Intuition behind \adv:}
How do these properties of real-world temporal hypergraphs provide efficiency to \adv? 
Here, we provide some reasons why we expect \adv to be faster and more space-efficient than \naive and \wsdmshort.

\begin{itemize}[leftmargin=*]
    \item \textbf{Connection to Obs. 5:} Each node in the projected graph $P$ used in \naive represents a unique temporal hyperedge, and its size heavily depends on $\delta$.
    On the other hand, the nodes in the projected graph $Q$ maintained in \adv represent induced hyperedges, and several temporal hyperedges can share the same node. 
    Thus, more repetitions of temporal hyperedges provide higher efficiency of \adv, as observed in real-world temporal hypergraphs. 
    \item \textbf{Connection to Obs. 6:} The benefits of temporal locality of temporal hyperedges are two-fold: 
    (1) The tendency of temporal hyperedges to repeat within a short period of time indicates that duplicated temporal hyperedges are more likely to co-appear in the temporal window in \adv, which reduces the size of the projected graph $Q$. 
    (2) If duplicated temporal hyperedges reappear within the temporal window, insertion/deletion of nodes and edges of $Q$ are skipped, which is beneficial in terms of speed.
\end{itemize}

    \section{Conclusion}
    \label{sec:conclusion}
    In this work, we propose (a) temporal hypergraph motifs (\motifs), which are tools for analyzing design principles of time-evolving hypergraphs, and (b)
\adv, which is a fast algorithm for exactly counting \motifs' instances.
Using them, we investigate $11$ real-world hypergraphs from $5$ domains. 
Our contributions are summarized as follows.

\begin{itemize}[leftmargin=*]
    \item \textbf{New concept:} We define $96$ temporal hypergraph motifs (\motifs) that describe local relational and temporal dynamics in time-evolving hypergraphs. 
    \item \textbf{Fast and exact algorithm:} We develop \adv, a fast and exact algorithm for counting the instances of \motifs. It is at most $2,163\times$ faster than the baseline approach.
    \item \textbf{Empirical discoveries:} \motifs reveal interesting structural and temporal patterns in real-world hypergraphs. \motifs also provide informative features that are useful in predicting future hyperedges.
\end{itemize}

\noindent \textbf{Reproducibility:} The source code and datasets used in this work are available at \url{https://github.com/geonlee0325/THyMe}.
    
    {\small \smallsection{Acknowledgements:} This work was supported by National Research Foundation of Korea (NRF) grant funded by the
Korea government (MSIT) (No. NRF-2020R1C1C1008296) and Institute of Information \& Communications
Technology Planning \& Evaluation (IITP) grant funded by the Korea government (MSIT) (No. 2019-0-00075, Artificial Intelligence Graduate School Program (KAIST)).}
    
	\bibliographystyle{IEEEtran}
	\bibliography{BIB/ref}

    \vspace{-1mm}
    \appendix
    \label{sec:appendix}
    \subsection{Details of \wsdm (\wsdmshort)}
\label{sec:appendix:wsdm}


The procedure \texttt{count}  (lines~\ref{alg:wsdm:count}-\ref{alg:wsdm:count:end}) counts the instances of \motifs that induce a set of $\ell$ connected static hyperedges.
That is, given a set $s=\{\tilde{e}_1,\dots,\tilde{e}_{\ell}\}$ of $\ell$ connected static hyperedges, \texttt{count} first constructs a time-sorted sequence $e(s)$ of temporal hyperedges whose nodes is one of $s$ (line~\ref{alg:wsdm:sort}).
It also introduces a map $C$ that maintains the counts of ordered hyperedges of length at most $\ell$.
Then \texttt{count} scans through the temporal hyperedges in $e(s)$ and tracks the subsequences that occur within the temporal window that spans temporal hyperedges within $\delta$ time units.
As the temporal window slides through the temporal hyperedges $e(s)$, the count of the sequences are computed based on the subsequences counted in $C$.
Refer to \cite{paranjape2017motifs} for more intuition behind this dynamic programming formulation.

\begin{algorithm}[h]
    \small
	\caption{\wsdmshort: Preliminary Algorithm for Exact Counting of \shorts' Instances\label{alg:wsdm}}
	\DontPrintSemicolon
	\SetKwInOut{Input}{Input}
    \SetKwInOut{Output}{Output}
    \SetKwFunction{Fcount}{count}
    \SetKwFunction{Fincrement}{increment}
    \SetKwFunction{Fdecrement}{decrement}
    \SetKwComment{Comment}{$\triangleright$}{}
    \Input{(1) temporal hypergraph: $T=(V,\mathcal{E})$\\(2) time interval $\delta$}
    \Output{\# of each temporal h-motif $t$'s instances: $M[t]$}
    \vspace{3pt}
    
    $S \leftarrow$ set of instances of static h-motifs in $G_T$\label{alg:wsdm:enum}\\
    \For{\upshape\textbf{each} instance $\{\tilde{e}_i,\tilde{e}_j,\tilde{e}_k\}\in S$}{
        \Fcount{$\{\tilde{e}_i,\tilde{e}_j,\tilde{e}_k\}$}\\
    }
    \For{\upshape\textbf{each} pair of overlapping hyperedges $\{\tilde{e}_i,\tilde{e}_j\}\in \wedge_{\mathcal{E}}$}{
        \Fcount{$\{\tilde{e}_i,\tilde{e}_j\}$}\\
    }
    \For{\upshape\textbf{each} hyperedge $\tilde{e}_i\in E_{\mathcal{E}}$}{
        \Fcount{$\{\tilde{e}_i\}$}\\
    }
    \Return{$M$}\\
    \vspace{3pt}
    
    \SetKwProg{myproc}{Procedure}{}{}
    \myproc{\Fcount{$s=\{\tilde{e}_1,\dots,\tilde{e}_\ell\}$}\label{alg:wsdm:count}}{
        $e(s)\leftarrow \text{sorted}(I(\tilde{e}_1) \cup \dotsi \cup I(\tilde{e}_\ell))$\label{alg:wsdm:sort}\\
        $w_s \leftarrow 1$\\
        $C \leftarrow$ map initialized to 0\\
        \For{\upshape\textbf{each} temporal hyperedge $e_i=(\tilde{e}_i,t_i)\in e(s)$}{
            \While{$t_{w_s} + \delta < t_i$}{
                \Fdecrement{$\tilde{e}_{w_s}$}\\
                $w_s \leftarrow w_s + 1$\\
            }
            \Fincrement{$\tilde{e}_i$}\\
        }
        \For{\upshape\textbf{each} $\langle e_i,e_j,e_k \rangle \in \text{permutations}(\{\tilde{e}_i,\tilde{e}_j,\tilde{e}_k\})$}{
            $M[h(\tilde{e}_i,\tilde{e}_j,\tilde{e}_k)]$ += $C[\text{concat}(\tilde{e}_i,\tilde{e}_j,\tilde{e}_k)]$\label{alg:wsdm:count:end}\\
        }
    }
    \vspace{2pt}
    \myproc{\Fincrement{$\tilde{e}$}}{
        \For{\upshape\textbf{each} prefix \textbf{in} $C$.keys.reverse of length $< \ell$}{
            $C[\text{concat}(\text{prefix},\tilde{e})]$ += $C[\text{prefix}]$\\
        }
        $C[\tilde{e}] \leftarrow C[\tilde{e}] + 1$
    }
    \vspace{2pt}
    \myproc{\Fdecrement{$\tilde{e}$}}{
        $C[\tilde{e}] \leftarrow C[\tilde{e}] - 1$\\
        \For{\upshape\textbf{each} suffix \textbf{in} $C$.keys of length $< \ell-1$}{
            $C[\text{concat}(\tilde{e},\text{suffix})]$ -= $C[\text{suffix}]$\\
        }
    }
\end{algorithm}

\vspace{-2mm}
\subsection{Details of Datasets~\label{sec:appendix:dataset}}
We provide the details of the eleven real-world temporal hypergraphs from the following five distinct domains:
\begin{itemize}[leftmargin=*]
    \item \textbf{email}: Each node is an email account and each hyperedge is the set of sender and receivers of the email.
    \item \textbf{contact}: Each node is a person and each hyperedge is a group interaction among people.
    \item \textbf{threads}: Each node is a user and each hyperedge is a group of users working in a thread.
    \item \textbf{tags}: Each node is a tag and each hyperedge is a set of tags attached to the question.
    \item \textbf{coauthorship}: Each node is an author and each hyperedge is a set of authors of the publication.
\end{itemize}

\end{document}